\documentclass[pre,amsmath,amssymb,twocolumn,superscriptaddress]{revtex4-2}
\usepackage{amsmath,amssymb}
\usepackage[usenames]{color}
\usepackage{amssymb}
\usepackage{dsfont}
\usepackage{grffile}
\usepackage{textcomp}
\usepackage{xspace}
\usepackage{hyperref}
\usepackage{graphicx} 
\usepackage{outlines}
\usepackage{natbib}

\newcommand{\tr}{{\rm tr}}

\newcommand{\im}{{\rm i}}

\newcommand{\sx}{\sigma^x}
\newcommand{\sy}{\sigma^y}
\newcommand{\sz}{\sigma^z}

\newcommand{\vlr}{v_{\rm LR}}

\newcommand{\sutd}{Science, Mathematics and Technology Cluster, Singapore
University of Technology and Design, 8 Somapah Road, 487372 Singapore}
\newcommand{\sutdepd}{EPD Pillar, Singapore University of Technology and Design, 8 Somapah Road, 487372 Singapore}
\newcommand{\cqt}{Centre for Quantum Technologies, National University of Singapore 117543, Singapore}
\newcommand{\majulab}{MajuLab, CNRS-UNS-NUS-NTU International Joint Research Unit UMI 3654, Singapore}

\begin{document}
\title{Relaxation exponents of OTOCs and overlap with local Hamiltonians}

\author{Vinitha Balachandran}
\affiliation{\sutd}
\author{Dario Poletti}
\affiliation{\sutd}
\affiliation{\sutdepd}
\affiliation{\majulab}
\affiliation{\cqt}
\affiliation{The Abdus Salam International Centre for Theoretical Physics, Strada Costiera 11, 34151 Trieste, Italy}

\begin{abstract}
  OTOC has been used to characterize the information scrambling in quantum systems. Recent studies showed that  local conserved quantities play a crucial role in governing the relaxation dynamics of OTOC in non-integrable systems. In particular, slow scrambling of OTOC is seen for observables that has an overlap with local conserved quantities.
  However, an observable may not overlap with the Hamiltonian, but with the Hamiltonian elevated to an exponent larger than one. Here, we show that higher exponents correspond to faster relaxation, although still algebraic, and with exponents that can increase indefinitely.
  Our analytical results are supported by numerical experiments.
  \end{abstract}
\maketitle

\setcounter{figure}{0}

\section{Introduction}
For generic many body quantum systems, information initially encoded in a few local degrees of freedom can spread in time over the entire space accessible. This process is called information scrambling and can be characterized by out of time ordered correlators (OTOCs)~\cite{Witten1998, Maldacena1999, HaydenPreskill2007, SekinoSussking2008, Shenker_2014, SachdevYe1993, Kitaev, LashkariHayden2013,  RobertsStanford2015, CotlerTezuka2017, RobertsSussking2015, HosurYoshida2016, Borgonovi2019,LiDu2017, GarttnerRey2017, LandsmanMonroe2019, Niknam2020, JoshiRoos2020, BlokSiddiqi2021, MiYu2021, Jochen2021}. For quantum systems with a classical limit, OTOCs can be mapped to Lyapunov exponents ~\cite{Galitski2017, Hashimoto_2017, Cotler_2018, Ignacio2018, Ch_vez_Carlos_2019, Fortes2019, Rammensee2018, Prakash2020, Bergamasco2019, Rozenbaum2020, Wang2020, wang2020quantum}. Because of this, OTOCs have been applied to understand the thermalization in many body quantum systems~\cite{HaydenPreskill2007, SekinoSussking2008, Shenker_2014, SachdevYe1993, Kitaev, LashkariHayden2013,  RobertsStanford2015, CotlerTezuka2017, RobertsSussking2015, HosurYoshida2016, Borgonovi2019}.

Recent studies have pointed out the relevance of local conserved quantities in the relaxation dynamics of OTOCs ~\cite{RakovszkyKeyserlingk2018, NahumHaah2017, NahumHaah2018, KeyserlingkSondhi2018, RakovszkyKeyserlingk2018, KhemaniHuse2018, Balachandran2021, Balachandran2022}.
In particular, in \cite{Balachandran2021} it was shown that the emergence of algebraic relaxation can stem from locality of the Hamiltonian, i.e., the ensuing presence of a Lieb-Robinson bound \cite{LiebRobinson1972}, and the eigestate thermalization hypothesis (ETH) \cite{Srednicki_1996, Deutsch}.
Importantly, with the approach developed in \cite{Balachandran2021} it was also possible to show that algebraic relaxation of the OTOC is typical.

In the scenarios considered until now, the operators in the OTOC would have non-zero overlap with the Hamiltonian or a local conserved quantity (i.e. total magnetization). Here we investigate how the relaxation dynamics would be affected if the operators in the OTOC, e.g. $A$, do not overlap with the Hamiltonian $H$, but only with one of its powers, i.e. $\tr(AH)=0$, but $\tr(AH^m) \ne 0$ for $m$ being an integer larger than one.
We show that depending on the exponent $m$ at which the overlap becomes non-zero, we expect an algebraic relaxation of the OTOC in time with an exponent proportional to $m$.
To obtain this result, we also show the relation between the first non-zero derivative of the diagonals of an operator in the energy basis, with the exponent $m$ at which $\tr(AH^m) \ne 0$.


The paper is organized as follows. In section \ref{sec:otoc} we introduce the definition of OTOCs as well as explain the relaxation dynamics of OTOCs from the knowledge of the matrix elements of the observables in the eigenenergy basis. In Sec.~\ref{sec:powerlaw} we show analytically our main result, i.e. that any different exponents can emerge in the relaxation of the OTOC, depending on the order at which the operators in the OTOC overlap with the Hamiltonian.
Our numerical results are presented in \ref{sec:results}. We draw our conclusions in Sec. \ref{sec:conclude}.

\section{Emergence of slow scrambling}\label{sec:otoc}
\subsection{Definition}
Consider the infinite-temperature out-of-time-ordered correlator (OTOC) between two local observables $A$ and $B$ defined as
\begin{align}\label{otoc}
 O^{AB}(t) &= \frac{1}{2}\langle [A(t), B][A(t), B]^\dagger\rangle
\end{align}
where $A(t) = U^{\dagger}A U$ is the time evolved operator $A$ due to the unitary evolution $U=\mathcal{T}e^{-\im \int_0^t H(\tau) d\tau}$ from the time-ordered integration of the (generically) time-dependent Hamiltonian $H(t)$. Expanding the commutators, we can rewrite Eq.(\ref{otoc}) as
\begin{align}\label{otoc1}
 \frac{1}{2}\langle [A(t), B][A(t), B]^\dagger\rangle &=\langle B^2A(t)^2\rangle -\langle A(t)BA(t)B\rangle \nonumber \\
 &=G^{AB}(t)-F^{AB}(t),\nonumber \\
\end{align}
where $G^{AB}(t)=\langle B^2A(t)^2\rangle$ is the time-ordered part of OTOC and $F^{AB}(t)=\langle A(t)BA(t)B\rangle$ is the not-time-ordered part. We consider only unitary and Hermitian observables for which $G(t)=1$ and hence we restrict ourselves to $F(t)$ in the remaining part.
Taking energy eigenstates as the basis of the Hilbert space, the time evolution of OTOC can be written in the eigenenergy basis $|p\rangle$ as
\begin{align}\label{otoc_energybasis}
 F^{AB}(t)= \frac{1}{\mathcal{V}}\sum_{p,q,k,l}e^{\im(E_p-E_q+E_k-E_l)t}A_{pq}B_{qk}A_{kl}B_{lp}
\end{align}
where $E_p$ is the eigenenergy,  $A_{pq}=\langle p|A|q \rangle$ and $B_{qk}=\langle q|B|k \rangle$. We work in units for which $\hbar=1$.

As $t\rightarrow\infty$, dominant terms in the above expression are those for which $E_p-E_q+E_k-E_l=0$. Hence, for generic systems \cite{Srednicki_1998, HuangZhang2019}, the infinite-time value of $F^{AB}(t)$ are given by
\begin{align}\label{eq:otocdiag}
 F^{AB}(\infty) = \frac{1}{\mathcal{V}} \bigg(\sum_{p}  &  A_{pp}^2B_{pp}^2   + \sum_{p,q\ne p}\big( A_{pp}B_{pq}A_{qq}B_{qp} \nonumber \\
       + &  A_{pq}B_{qq}A_{qp}B_{pp}\big)\bigg).
\end{align}
Eq. (\ref{eq:otocdiag}) highlights the importance of diagonal elements of $A$ and $B$ in the eigenenergy basis in the infinite-time value of OTOC. Indeed, a non-zero diagonal element in $A$ or $B$ is necessary to guarantee a non-zero value of $F^{AB}(\infty)$.

\subsection{Conditions for algebraic relaxation of OTOC}\label{sec:condition}
Two sufficient conditions for the emergence of algebraic relaxation of OTOC \cite{Balachandran2021,Balachandran2022} are
\begin{itemize}
  \item a Lieb-Robinson bound (or even an algebraic
spreading of correlation which occurs in systems with
power-law interactions),
  \item algebraic scaling of
infinite-time value of the OTOC with the system size.
\end{itemize}
In local and bounded Hamiltonians, the speed of propagation of the correlations is limited by Lieb-Robinson bound \cite{LiebRobinson1972, CheneauKuhr2012}. Hence, an accurate description of the evolution of OTOC of a thermodynamically large system, can be obtained simply considering a finite portion of it. Assuming that the system is maximally scrambled within the region of size $L$, the decay of $F^{AB}_{L=\infty}(t)$ is bounded by the Lieb-Robinson velocity $\vlr$ as
\begin{align}
F^{AB}_{L=\infty}(t) \approx F^{AB}_{L=s\;\vlr \;t}(\infty), \label{eq:inf_time}
\end{align}
 where $s$ is a real number larger than $1$.
Hence, $L$ increases with time and is a time-dependent quantity.
Therefore, the scaling of $F^{AB}_{L}(\infty)$ is crucial to predict the bound for the relaxation of OTOC. In particular, when  $F^{AB}_{L}(\infty)$ decays algebraically with the system size, e.g. $F^{AB}_{L}(\infty)\propto L^{-\alpha}$, then the OTOC of the thermodynamic size system cannot decay faster than algebraically in time, or more precisely from Eq.(\ref{eq:inf_time}) one can write that it cannot be faster than
\begin{align}
F^{AB}_{L=\infty}(t) \propto \frac{1}{t^\alpha} \label{eq:slow_decay}
\end{align}
because $L=s\;\vlr \;t$.

The actual decay of the OTOC may even be slower, for example considering cases in which the system goes through prethermalization \cite{LuitzKhemani} or in which the system is many-body localized \cite{Lee2019}. However, the relaxation cannot be faster, hence the OTOC will have a slow, non-exponential relaxation. A comprehensive analysis of this is presented in \cite{Balachandran2021}.

\section{Generic algebraic relaxation in short-ranged systems }\label{sec:powerlaw}

\subsection{Estimate of the infinite time value of OTOC}
In this section, we show how to obtain the approximate value of the infinite-time, finite-size, OTOC $F^{AB}_{L}(\infty)$
\begin{align}
 F^{AB}_{L}(\infty) = & \frac{1}{\mathcal{V}}  \sum_{p} A_{pp}^2B_{pp}^2  + \frac{1}{\mathcal{V}}  \sum_{p,q\ne p} A_{pp}A_{qq}|B_{pq}|^2 \nonumber \\
  & + \frac{1}{\mathcal{V}}  \sum_{p,q\ne p} B_{pp}B_{qq}|A_{pq}|^2 \nonumber \\
  \approx    & \frac{1}{\mathcal{V}}  \sum_{p} A_{pp}^2B_{pp}^2  + \frac{1}{\mathcal{V}}\sum_{ p} A_{pp}A_{pp}\left[(B B^{\dagger})_{pp}-  B_{pp}^2\right]\nonumber \\
  & + \frac{1}{\mathcal{V}}\sum_{ p} B_{pp}B_{pp}\left[(A A^{\dagger})_{pp}-  A_{pp}^2\right]\nonumber \\
  \approx   & \frac{1}{\mathcal{V}}  \sum_{p} A_{pp}^2B_{pp}^2  + \frac{1 }{\mathcal{V}} \sum_{p} \left[ \tr{(B B^{\dagger})}- B_{pp}^2\right] A_{pp}^2  \nonumber \\
  & + \frac{1 }{\mathcal{V}} \sum_{p} \left[ \tr{(A A^{\dagger})}- A_{pp}^2\right] B_{pp}^2  \nonumber \\
  \approx   & \frac{1}{\mathcal{V}}  \sum_{p} A_{pp}^2B_{pp}^2  + \frac{1 }{\mathcal{V}} \sum_{p} \left[ 1- B_{pp}^2\right] A_{pp}^2 \nonumber \\
  & + \frac{1 }{\mathcal{V}} \sum_{p} \left[ 1- A_{pp}^2\right] B_{pp}^2 \nonumber \\
  \approx   & \frac{1 }{\mathcal{V}} \sum_{p} \left[A_{pp}^2+B_{pp}^2  - A_{pp}^2 B_{pp}^2\right] \nonumber \\
  \approx   & \frac{1 }{\mathcal{V}} \sum_{p} \left[A_{pp}^2+B_{pp}^2 \right],    \label{eq:otocdiagof}
\end{align}
where we have used steps similar to \cite{HuangZhang2019, Balachandran2021}, and a similar discussion can be found in \cite{Balachandran2022}.
Thus, the main contribution of the infinite-time finite-size OTOC comes from the $A_{pp}^2$ and $B_{pp}^2$ terms which we will be discussing in the following.

\subsection{Structure of the diagonal elements}\label{sec:structure}
In short, the diagonal element $A_{pp}$ can be approximated by a function of eigenenergy $E_p$
\begin{align}\label{eq:lemma2a}
 |A_{pp}- f_A(E_p/L)| &  \leq e^{-(\Omega(L))},
\end{align}
where $f_A(E_p/L)$ can be expanded as
\begin{align}\label{eq:lemma2b}
 f_A(E_p/L) & = f_A(0)+f^{(1)}_A(0)E_p/L+\frac{1}{2}f^{(2)}_A(0)E_p^2/L^2+... \nonumber \\
 & = \sum_q \frac{f_A^{(q)}}{q!} \left(\frac{E_p}{L}\right)^q
\end{align}
with $f_A^{(q)}$ being the $q-$th derivative of $f_A$.
We also note that, using Lemma 1 in \cite{HuangZhang2019} one can write
\begin{align}\label{eq:lemma1}
 \frac{1}{\mathcal{V}}\sum_p E_p^q & = \langle H^q \rangle = O(L^{q/2}).
\end{align}
In \cite{HuangZhang2019} it was shown that, for traceless operators $f_A(0)=0$, and if $f^{(1)}_A(0)\ne 0$ then we can write
\begin{align}\label{eq:overlap1}
\tr{(A H)} & =  \frac{1}{\mathcal{V}}  \sum_{p}  A_{pp} E_p \nonumber \\
& \approx\frac{1}{\mathcal{V}L}  \sum_{p}E_p^2 f^{(1)}_A(0) \nonumber \\
& \approx\frac{\langle H^2 \rangle}{L}   f^{(1)}_A(0)
\end{align}
and thus
\begin{align}\label{eq:derv1}
 f^{(1)}_A(0) & \approx \frac{\tr{(A H)} L }{\langle H^2 \rangle}.
\end{align}
Hence, the first derivative of a local observable $A$ is independent of the system size.
From Eq,~(\ref{eq:derv1}) we get
\begin{align}\label{eq:term1a}
  F_L^{AB}(\infty)& \approx \frac{1}{\mathcal{V}} \sum_{p}    (A_{pp}^2+B_{pp}^2) \nonumber \\
  &\approx  \frac{1}{\mathcal{V}}\sum_{p}\frac{E_p^2}{L^2} \left[\left(f^{(1)}_A(0)\right)^2 + \left(f^{(1)}_B(0)\right)^2 \right]   \nonumber \\
  &\approx  \frac{1}{\mathcal{V}}\sum_{p}\frac{E_p^2}{L^2}\frac{\left[\tr{(A H)}^2 +\tr{(B H)}^2\right]L^2 }{\langle H^2 \rangle^2}  \nonumber \\
  &\approx  \frac{\tr{(A H)}^2 +\tr{(B H)}^2}{\langle H^2 \rangle} \nonumber \\
  &\propto \frac 1 L.
\end{align}
The last step stems from the fact that $\tr(AH)$ and $\tr(BH)$ are independent of the system size, while $\langle H^2 \rangle \propto L$ from Eq.~(\ref{eq:lemma1}).

If $\tr(AH)= 0$ but, for instance, $\tr(AH^p) \neq 0$ only for $p\ge p_c$ then one can generalize the previous result.
Considering $f_A^{(q)}(0)$ as the smalles non-zero derivative of $f_A$ at zero energy (with the same parity as $p_c$), then we can write
\begin{align}
    \tr(AH^{p_c}) &= \sum_n \frac{f_A^{(q)}}{q!}\frac{E_n^{q+p_c}}{L^q}
\end{align}
which implies that
\begin{align}\label{eq:fapprox}
    f_A^{(q)} &= \frac{q! \tr(AH^{p_c}) L^q}{\langle H^{p_c+q} \rangle}.
\end{align}
Now, if $q<p_c$ then $f_A^{(q)}$ would decay as $L^{-(p_c-q)/2}$, which implies that they are $0$, and the non-size dependent $f_A^{(q)}(0)$ would occur exactly at $q=p_c$. This implies that the first non-zero derivative of $f_A(0)$ is the $p_c-$th one.
Thus, when $\tr(AH^p) \neq 0$ only for $p\ge p_c$ we can write
\begin{align}\label{eq:termma}
  F_L^{AB}(\infty)& \approx  \frac{1}{\mathcal{V}}\sum_{n} \left( \frac{E_n}{L} \right)^{2p_c} \left[\left(f^{(p_c)}_A(0)\right)^2 + \left(f^{(p_c)}_B(0)\right)^2 \right]   \nonumber \\
  &\approx  \frac{1}{\mathcal{V}}\sum_{n} \left( \frac{E_n}{L} \right)^{2p_c}\frac{\left[\tr{(A H^{p_c})}^2 +\tr{(B H^{p_c})}^2\right]L^{2p_c} }{\langle H^{2p_c} \rangle^2}  \nonumber \\
  &\approx  \frac{\tr{(A H^{p_c})}^2 +\tr{(B H^{p_c})}^2}{\langle H^{2p_c} \rangle} \nonumber \\
  &\propto \frac 1 {L^{p_c}}.
\end{align}
Building on Eq.(\ref{eq:termma}), and combining it with the Lieb-Robinson bound $L=s\;\vlr \;t$ we can thus guarantee that $F^{AB}$ cannot relax faster than $t^{-p_c}$. Furthermore, for systems in which correlations mostly spread diffusively, i.e. proportional to $t^{1/2}$, we can can expect $F^{AB}$ to relax as $t^{-p_c/2}$.
%
Hence, the structure of the diagonal elements of the observables, and which is the first non-zero derivative at $0$ energy, i.e., which is the first exponent of the Hamiltonian that has non-zero overlap with the operators $A$ and $B$ considered, play an important role in the relaxation dynamics of the OTOC in the system. This is numerically verified in detail in the following section.


\section{Results}\label{sec:results}
\subsection{Model}
 We consider a prototypical non-integrable model, the tilted Ising chain with Hamiltonian
\begin{align}
H & = \sum_{l=1}^{L-1}J_z \sz_l\sz_{l+1} + \sum_{l=1}^{L} (h_x \sx_l+ h_z \sz_l),       \label{eq:Ising}
\end{align}
where $J_z$ is the coupling constant in the z direction, while $h_x$ and $h_z$ are the transverse and the longitudinal field strengths. The model is integrable when either $h_x=0$ or $h_z=0$. This can be verified by studying the level spacing statistics which typically follows a Poisson distribution for integrable systems and a Wigner-Dyson distribution for non-integrable
ones \cite{BGS1984,Casati1980}. In particular, one can calculate $\delta_{n}=E_{n+1}-E_n$, the level spacing between two consecutive energy levels $E_n$ and $E_{n+1}$ within a single symmetry sector, define the ratio $r_n = \text{max}(\delta_n, \delta_{n+1})/ \text{min}(\delta_n, \delta_{n+1})$ and take an average $r =\sum_n r_n/N$ where N is the number of energy level differences considered. For a Poisson distribution, $r$ can be computed analytically and it gives $r=2 \ln 2 - 1 \approx 0.386$, while for a Wigner-Dyson distribution $r$ can be evaluated numerically to be $r \approx 0.529$ \cite{OganesyanHuse2007}. In the current work, we use parameters $J_z = 1$, $h_z=0.809$ and $h_x=0.9$ which result in $r \approx 0.53$ already for a system size of $L=12$ spins.

\subsection{Observables and structure of their diagonal elements}
\begin{figure}
\includegraphics[width=\columnwidth]{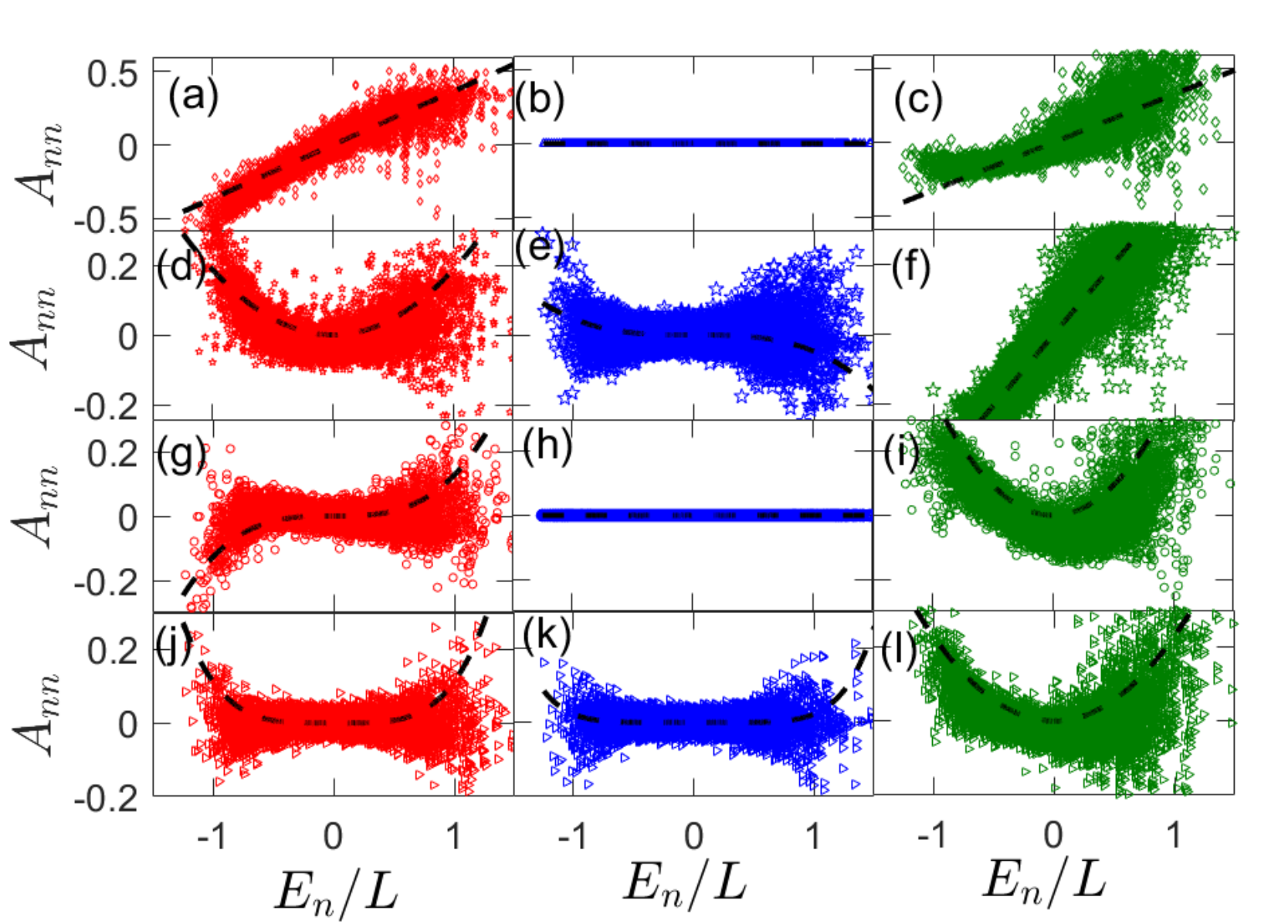}
\caption{Diagonal elements of the observable in the energyeigen basis for single site observables $A=\sigma_{L/2}^\alpha$ panel(a-c), double site observables $A=\sigma_{L/2}^\alpha\sigma_{L/2+1}^\alpha$ (d-f), triple site observables $A=\sigma_{L/2-1}^\alpha\sigma_{L/2}^\alpha\sigma_{L/2+1}^\alpha$ (g-i) and quadruple site observables $A=\sigma_{L/2-2}^\alpha\sigma_{L/2-1}^\alpha\sigma_{L/2}^\alpha\sigma_{L/2+1}^\alpha$ (j-l). Left panels are for $\sx_l$ ($\alpha=x$) observables, middle panels are for $\sy_l$ ($\alpha=y$) observables and right panels are for $\sz_l$ ($\alpha=z$) observables. Dashed lines are the lowest order fits in the Taylor expansion of the observable in Eq. (\ref{eq:lemma2b}). Here, $L=14$, $J_z=1$, $h_x=0.9$ and $h_z=0.809$.}
\label{fig:Fig1r}

\end{figure}

To span over a variety of different structures, and to have operators $A$ which have $\tr(A H^p)\ne 0$ only for $p\le p_c$ with $p_c$ which can be different from $1$, we analyze both single-site and multi-site observables in our study. In particular, we consider the following four types of observables,
\begin{align}
  {\rm single-site} \rightarrow & \;\; \sigma_l^{\alpha} \label{eq:single-site}\\
  {\rm double-site} \rightarrow & \;\; \sigma_l^{\alpha}\sigma_{l+1}^{\alpha} \label{eq:double-site}\\
  {\rm triple-site} \rightarrow & \;\; \sigma_{l-1}^{\alpha}\sigma_l^{\alpha}\sigma_{l+1}^{\alpha} \label{eq:triple-site}\\
  {\rm quadruple-site} \rightarrow & \;\; \sigma_{l-2}^{\alpha}\sigma_{l-1}^{\alpha}\sigma_{l}^{\alpha}\sigma_{l+1}^{\alpha} \label{eq:quadruple-site}
\end{align}
where $\alpha = x,y$ or $z$. The diagonal elements of these operators in the eigenbasis of Hamiltonian Eq.~(\ref{eq:Ising}) are shown in Fig.~\ref{fig:Fig1r}. The left column is for $\alpha=x$, center column for $\alpha=y$ and right column for $\alpha=z$. The rows are for increasing range of operators from top to bottom, with the top row for single-site operators and the fourth row for four-site operators. In all the panels, the dashed lines represent the expected algebraic energy dependence of $f_A$ near energy zero from Sec.~\ref{sec:structure}. We note that these fits are evaluated directly from calculating $f^{(n)}_A(0)$ with Eq.(\ref{eq:fapprox}) along with eigenenergies $E_n$ for the system Hamiltonian in Eq.~(\ref{eq:Ising}) with no fitting parameters.

For single site observables $A=\sx_l$ and $\sz_l$, for the non-integrable Ising chain $\tr(AH)\neq0$ and so $f^{(1)}_A(0)\neq0$. However, with $A=\sy_l$, $\tr(AH^n)=0$ for any $n$. Hence, we expect a linear variation of the diagonal elements of $\sx_l$ and $\sz_l$ with energy density $E_n/L$ and a flat profile for $\sy_l$. This can be seen in Fig. \ref{fig:Fig1r}(a-c). To conform our analytical predictions, we plot $f^{(1)}_A(0)E_n/L$ where $f^{(1)}_A(0)$ is calculated explicity form Eq.\ref{eq:derv1}.

The two-site observables $A=\sigma^\alpha_{L/2}\sigma^\alpha_{L/2+1}$, are shown in the panels (d-f).  For $A=\sx_{L/2}\sx_{L/2+1}$, $\tr(AH^2)\neq0$ whereas $\tr(AH)=0$ and, as predicted in Sec.~\ref{sec:structure}, we thus observe that $f_A$ can be fitted by a parabola $f^{(2)}_A(0)E_n^2/(2!L^2)$ indicated by the dashed black lines. Since $\tr(AH)\neq0$ for  $A=\sz_{L/2}\sz_{L/2+1}$, we see a linear scaling of $A_{nn}$ with $E_n/L$.  For $A=\sy_{L/2}\sy_{L/2+1}$ observable, $\tr(AH^m)\neq0$ for $m \ge 3$. Hence, we see a cubic structure of the diagonal elements with a fitting of the form $f^{(3)}_A(0)E_n^3/(3!L^3)$.

We also consider triple-site observables $A=\sigma^\alpha_{L/2-1}\sigma^\alpha_{L/2}\sigma^\alpha_{L/2+1}$.
These are depicted in the panels (g-i). Here, $\tr(AH^m)\neq0$ for $m \ge 3$ for $\sx_l$ observables and we clearly see a cubic structure for the diagonal elements that can be fitted with lines of the form $f^{(3)}_A(0)E_n^3/(3!L^3)$. Since there are no diagonal elements for any power of $H$ for the $\sy$ observable, a flat profile is seen. With the $\sz$ observable, a parabolic structure is seen since $\tr(AH^2)\neq0$ whereas $\tr(AH)=0$. This is also nicely fitted by $f^{(2)}_A(0)E_n^2/(2!L^2)$ in the panel (i).

For the four site observable we study $A=\sigma^\alpha_{L/2-2}\sigma^\alpha_{L/2-1}\sigma^\alpha_{L/2}\sigma^\alpha_{L/2+1}$. For $A=\sx_{L/2-2}\sx_{L/2-1}\sx_{L/2}\sx_{L/2+1}$, a quartic structure can be seen as $\tr(AH^m)\neq0$ only for $m \ge 4$. This is fitted by $f^{(4)}_A(0)E_n^4/(4!L^4)$ (black dashed lines). With $\sy_l$ observables, the expected structure is hexic (polynomial of sixth degree) because $\tr(AH^m)\neq0$ only for $m \ge 6$. Though it is less clear, we fit it with the expected scaling $f^{(6)}_A(0)E_n^6/(6!L^6)$ using the black dashed lines. For $\sz$ observable, we find a parabolic structure in accordance with our prediction as $\tr(AH^2)\neq0$ whereas $\tr(AH)=0$. Since we consider systems of size $L=14$, the results in Fig.~\ref{fig:Fig1r}(j-l) are partially affected by finite-size effects. Despite this, the numerics are aligned with our theoretical predictions.

To summarize this section, we observe clearly that the diagonal elements of operators can have a very different dependence as a function of energy near zero. In particular we have numerically verified the prediction that $A_{nn}\sim 1/L^{p}$ where $p$ is the lowest positive integer such that $\tr(AH^p)\ne 0$.

\subsection{Scaling of the infinite time value of OTOC}

\begin{figure}
\includegraphics[width=\columnwidth]{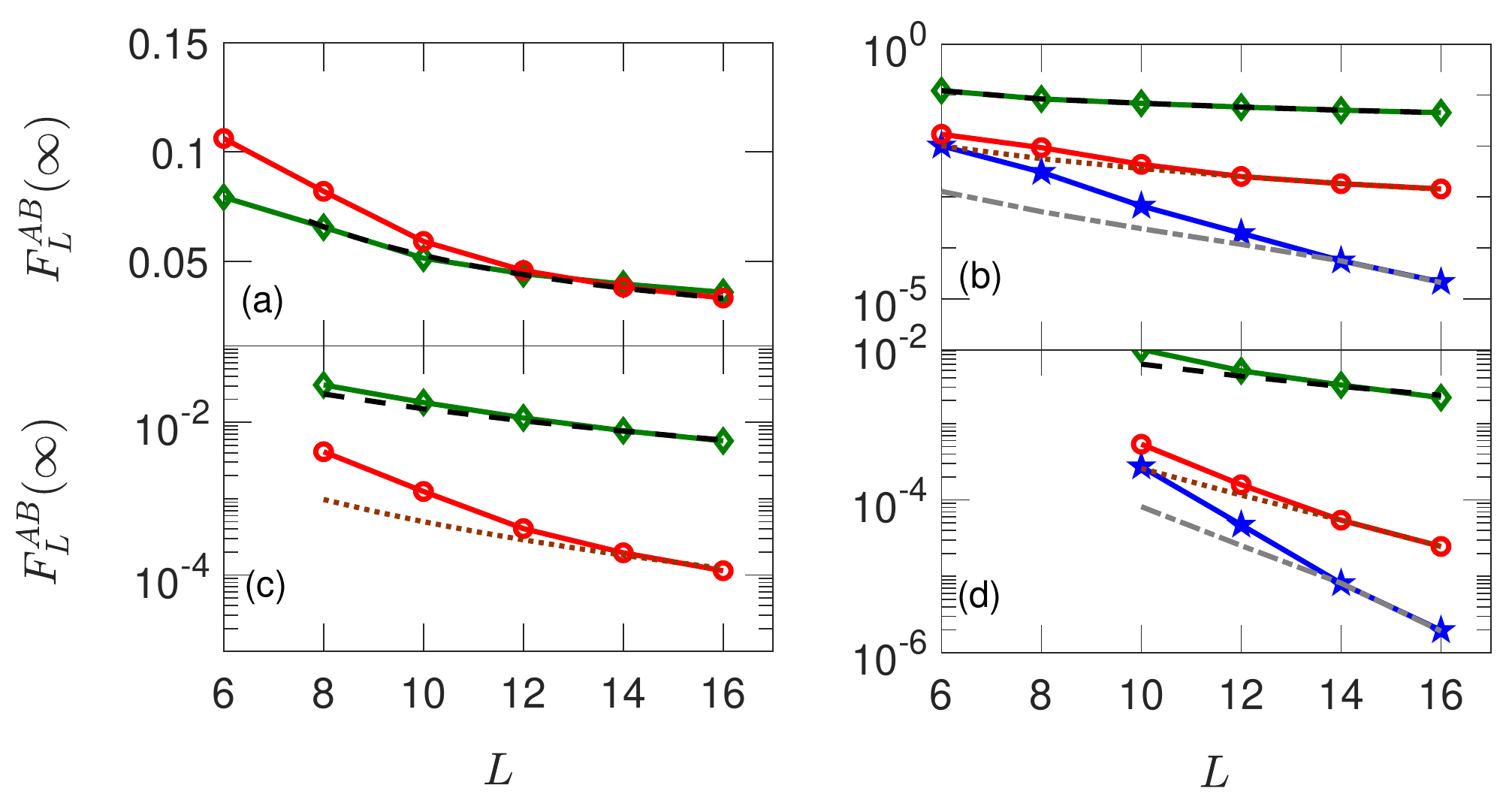}
\caption{Infinite time values of OTOC corresponding to the single site observables with $A=\sigma_{L/2-1}^\alpha$, $B=\sigma_{L/2}^\alpha$ panel(a), double site observables with $A=\sigma_{L/2-2}^\alpha\sigma_{L/2-1}^\alpha$, $B=\sigma_{L/2}^\alpha\sigma_{L/2+1}^\alpha$ (b), triple site observables with $A=\sigma_{L/2-3}^\alpha\sigma_{L/2-2}^\alpha\sigma_{L/2-1}^\alpha$,  $B=\sigma_{L/2}^\alpha\sigma_{L/2+1}^\alpha\sigma_{L/2+2}^\alpha$  (c) and quadruple site observables with $A=\sigma_{L/2-4}^\alpha\sigma_{L/2-3}^\alpha\sigma_{L/2-2}^\alpha\sigma_{L/2-1}^\alpha$, $B=\sigma_{L/2}^\alpha\sigma_{L/2+1}^\alpha\sigma_{L/2+2}^\alpha\sigma_{L/2+3}^\alpha$ (d). Green lines with diamonds are for observables involving only $\sz_l$ $(\alpha=z)$ operators, red lines with circles for $\sx_l$ $(\alpha=x)$ and blue lines with stars for $\sy_l$ $(\alpha=y)$ operators respectively. Black-dashed, brown-dotted and grey-dashed dotted lines are the fits for $\sz_l$, $\sx_l$ and $\sy_l$ observables. Here, $J_z=1$, $h_x=0.9$ and $h_z=0.809$.
}
\label{fig:Fig2r}
\end{figure}

%
%
In Fig.~\ref{fig:Fig2r} we show numerical confirmation that given the minimum positive integer $p_c$ such that $\tr( A H^{p_c})\ne 0$ or $\tr( B H^{p_c})\ne 0$, then $F^{AB}_{L=\infty}(t) \propto 1/L^{p_c}$.
In each of the panels we show how the infinite time value of the OTOC $F^{AB}_L(t=\infty)$ varies as a function of the system size $L$. In the different panels we will focus on single-site, panel (a), two-site, panel (b), three-site, panel (c) and four-site, panel (d), observables.
In each panel the red line with circles corresponds to $\alpha=x$, blue with stars to $\alpha=y$ and green with diamonds to $\alpha=z$.
In panel Fig.~\ref{fig:Fig2r}(a), we plot the infinite time values of OTOC with single site observables $A,B=\sigma^\alpha_{l}$, where $l=L/2$ for observable $B$ and $l=L/2-1$ for observable $A$. We see that these observables have $p_c=1$, and hence they follow $1/L$ scaling as shown by dashed line. $\sy_l$ has no overlap with any local conserved quantities and hence the diagonal elements as well as the infinite time values of OTOC are zero.
Fig.~\ref{fig:Fig2r}(b) is for double site observables Eq.~(\ref{eq:double-site}), where $l=L/2$ for observable $B$ and $l=L/2-2$ for observable $A$. We compare the numerical results with fitted lines, in particular with $1/L^2$ (dotted), $1/L^3$ (dashed dotted lines) and $1/L$ (dashed), respectively corresponding to operators with $p_c=2,3$ and $1$. We note that due to the small value of the overlap of $\sigma^y_{j}\sigma^y_{j+1}$ with the Hamiltonian, the expected scaling is followed only at larger system sizes.
%
In panel (c), we plot the triple site observables  Eq.~(\ref{eq:triple-site}) where $j=L/2-3$ for $A$ and $j=L/2$ for $B$. Fitted lines are for $1/L^2$ and $1/L^3$ scalings as expected since $\alpha=2$ and $3$ respectively. Since the diagonal elements of $\sigma^y_{j}\sigma^y_{j+1}\sigma^y_{j+2}$ are zero, the infinite time value of the OTOC $F^{AB}_L(t=\infty)$ is zero.
Panel (d) is for quadruple site observables  Eq.~(\ref{eq:quadruple-site}) where $j=L/2-4$ for $A$ and $j=L/2$ for $B$. The expected scalings are $p_c=4,6,2$. However, due to the fact that the observables have a large support at initial time, we see that the correct scaling of $1/L^4,1/L^6,1/L^2$ are followed only at large system sizes.


\subsection{Dynamics of OTOCs}
\begin{figure}
\includegraphics[width=\columnwidth]{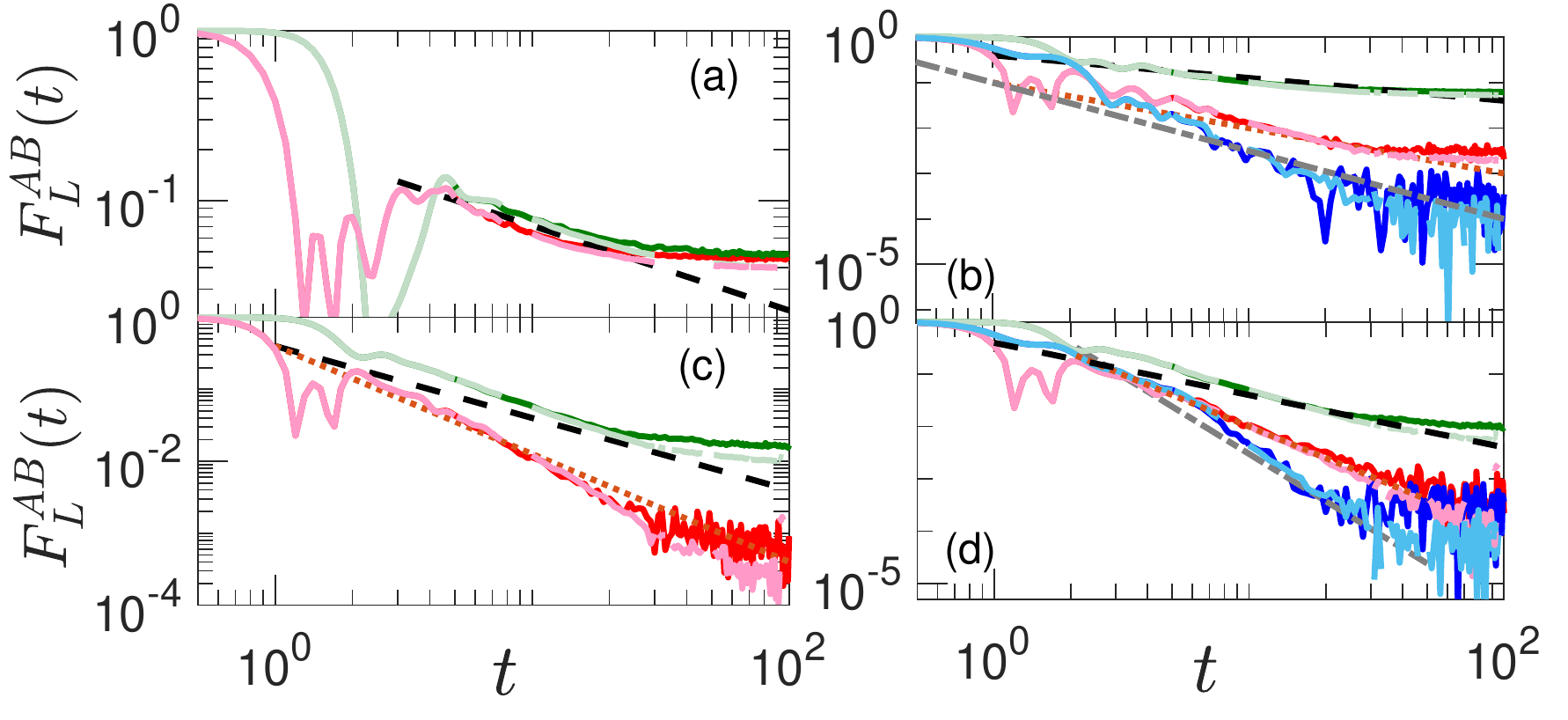}
\caption{Time evolution of OTOC corresponding to the single site observables with $A=\sigma_{L/2-1}^\alpha$, $B=\sigma_{L/2}^\alpha$ panel(a), double site observables with $A=\sigma_{L/2-2}^\alpha\sigma_{L/2-1}^\alpha$, $B=\sigma_{L/2}^\alpha\sigma_{L/2+1}^\alpha$ (b), triple site observables $A=\sigma_{L/2-3}^\alpha\sigma_{L/2-2}^\alpha\sigma_{L/2-1}^\alpha$,  $B=\sigma_{L/2}^\alpha\sigma_{L/2+1}^\alpha\sigma_{L/2+2}^\alpha$  (c) and quadruple site observables $A=\sigma_{L/2-4}^\alpha\sigma_{L/2-3}^\alpha\sigma_{L/2-2}^\alpha\sigma_{L/2-1}^\alpha$, $B=\sigma_{L/2}^\alpha\sigma_{L/2+1}^\alpha\sigma_{L/2+2}^\alpha\sigma_{L/2+3}^\alpha$  (d).Green lines are for observables involving only $\sz_l$ $(\alpha=z)$ operators, red lines for $\sx_l$ $(\alpha=x)$ and blue lines for $\sy_l$ $(\alpha=y)$ operators respectively. Black-dashed, brown-dotted and grey-dashed dotted lines are the fits for $\sz_l$, $\sx_l$ and $\sy_l$ observables discussed in the text. Here, $L=14$ for lighter shades and $L=12$ for darker shades and $J_z=1$, $h_x=0.9$ and $h_z=0.809$.
}
\label{fig:Fig3r}
\end{figure}
We study the dynamics of OTOC in Fig. ~\ref{fig:Fig3r}, where each panel reflects the same case analyzed in the corresponding panel of Fig.~\ref{fig:Fig2r}. Green lines are for observables involving only $\sz_l$ operators, red lines for $\sx_l$ and blue lines for $\sy_l$ operators respectively.
In these plots we need to study the long-time evolution. We thus need to disregard initial transients. At the same time, though, our results are affected by finite size, so we would need to concentrate on long, yet, intermediate times to evaluate the relaxation of the OTOC over time.
Light shades are for $L=14$ and dark shades for $L=12$. 
Black-dashed, brown-dotted and grey-dashed dotted lines are the fits for $\sz_l$, $\sx_l$ and $\sy_l$ observables.
Fig.~\ref{fig:Fig3r}(a) is for single-site observables, as in Eq.~(\ref{eq:single-site}). We have already seen that since $\tr(OH)\neq0$, for $ (O=A,B)$, then the infinite time value of OTOC $F^{AB}_{L}(t=\infty)$ scales as $1/L$. From our discussion at the end of Sec.~\ref{sec:structure}, we thus expect that $F^{AB}(t)\propto1/t^{1/2}$, and the numerical result of the dynamics is well fitted by the black dashed line proportional to $t^{1/2}$.

In Fig.~\ref{fig:Fig3r}(b) we study the two-site observables of Eq.~(\ref{eq:double-site}) with $l=L/2$ for observable $B$ and $l=L/2-2$ for observable $A$. As already discussed, the lowest order terms that has non zero values in the Taylor expansion for these observables are $1,2$ and $3$ respectively for $\sz_l,\sx_l,\sy_l$ observables. In Fig.~\ref{fig:Fig2r}(b) we showed the scaling of the infinite-time OTOC for these observables as $1/L$, $1/L^2$, $1/L^3$. Here we would thus expect a scaling with time of $1/t^{1/2}$, $1/t$ and $1/t^{3/2}$ which is shown in the plots by dotted, dashed dotted and dashed lines.
%
We study the evolution of three-site observables of Eq.~(\ref{eq:triple-site}) in Fig. \ref{fig:Fig3r}(c). Here $l=L/2-3$ for $A$ and $l=L/2$ for $B$. Fitted lines are for $1/t$ and $1/t^{3/2}$ scaling as expected since $\tr(OH)\neq0$ for $\sz_l$ and $\tr(OH^2)\neq0$ for $\sx_l$ observable.
Panel (d) shows the dynamics for four site observables with $l=L/2-4$ for $A$ and $l=L/2$ for $B$. The expected scaling is $1/t$, $1/t^{2}$ and $1/t^3$ for $\alpha = z,x$ and $y$ respectively, for which operators the corresponding critical exponent $p_c$ that give non-zero overlap are $2, 4$ and $6$.

\section{Conclusions}\label{sec:conclude}
OTOCs have been studied as a probe for quantum information scrambling. Slow, algebraic scrambling has been reported in systems with local conserved quantities \cite{Balachandran2021, RakovszkyKeyserlingk2018, KhemaniHuse2018, Balachandran2022}.

In this paper we showed that higher the exponent at which ones elevates the Hamiltonian in order to have a non-zero overlap with the operators in the OTOC, the faster is the relaxation of the OTOC over time. Furthermore, if there is an exponent such that the overlap is non-zero, then the relaxation, even if it appears to be fast, is bounded to be, at the fastest, algebraic, and only if there is no overlap with any power of the Hamiltonian (or other conserved quantities), then the relaxation can be exponential.

From our resuls it follows that considering single-site operators in the OTOC, and a local Hamiltonian with only single site and nearest neighbours term, relaxation can only take a limited set of exponents. It is thus necessary to consider operators with larger support, such as two-site, three-site and four-site operators, to observe a larger variety and magnitude of relaxation exponents. This, however, leads to the difficulty of study the relaxation numerically due to more pronounced finite-size effects when studying operators with larger support. Future developments in numerical methods could help to test our results for larger systems.

In order to derive these results, we also found a relation between the first non-zero derivative of the function representing the diagonals of an operator in the energy basis, and the first non-zero exponent of the Hamiltonian which has non-zero overlap with the operators of the OTOC.
Future works could extend these results to time-dependent systems with other types of conserved quantities.


{\it Acknowledgments:} We acknowledge C. von Keyserlingk who asked a critical and insightful question which started this project. 
The computational work for this article were partially performed on the National Supercomputing Centre, Singapore \cite{NSCC}. DP and VB are extremely grateful to G. Casati, to whom this volume is dedicated, for his guidance and example throughout the many years we had the fortune to learn from, interact and collaborate with.

\bibliography{bibliography}

\begin{thebibliography}{52}%
\makeatletter
\providecommand \@ifxundefined [1]{%
 \@ifx{#1\undefined}
}%
\providecommand \@ifnum [1]{%
 \ifnum #1\expandafter \@firstoftwo
 \else \expandafter \@secondoftwo
 \fi
}%
\providecommand \@ifx [1]{%
 \ifx #1\expandafter \@firstoftwo
 \else \expandafter \@secondoftwo
 \fi
}%
\providecommand \natexlab [1]{#1}%
\providecommand \enquote  [1]{``#1''}%
\providecommand \bibnamefont  [1]{#1}%
\providecommand \bibfnamefont [1]{#1}%
\providecommand \citenamefont [1]{#1}%
\providecommand \href@noop [0]{\@secondoftwo}%
\providecommand \href [0]{\begingroup \@sanitize@url \@href}%
\providecommand \@href[1]{\@@startlink{#1}\@@href}%
\providecommand \@@href[1]{\endgroup#1\@@endlink}%
\providecommand \@sanitize@url [0]{\catcode `\\12\catcode `\$12\catcode
  `\&12\catcode `\#12\catcode `\^12\catcode `\_12\catcode `\%12\relax}%
\providecommand \@@startlink[1]{}%
\providecommand \@@endlink[0]{}%
\providecommand \url  [0]{\begingroup\@sanitize@url \@url }%
\providecommand \@url [1]{\endgroup\@href {#1}{\urlprefix }}%
\providecommand \urlprefix  [0]{URL }%
\providecommand \Eprint [0]{\href }%
\providecommand \doibase [0]{https://doi.org/}%
\providecommand \selectlanguage [0]{\@gobble}%
\providecommand \bibinfo  [0]{\@secondoftwo}%
\providecommand \bibfield  [0]{\@secondoftwo}%
\providecommand \translation [1]{[#1]}%
\providecommand \BibitemOpen [0]{}%
\providecommand \bibitemStop [0]{}%
\providecommand \bibitemNoStop [0]{.\EOS\space}%
\providecommand \EOS [0]{\spacefactor3000\relax}%
\providecommand \BibitemShut  [1]{\csname bibitem#1\endcsname}%
\let\auto@bib@innerbib\@empty
\bibitem [{\citenamefont {Witten}(1998)}]{Witten1998}%
  \BibitemOpen
  \bibfield  {author} {\bibinfo {author} {\bibfnamefont {E.}~\bibnamefont
  {Witten}},\ }\bibfield  {title} {\bibinfo {title} {{Anti-de Sitter space and
  holography}},\ }\href {https://doi.org/10.4310/ATMP.1998.v2.n2.a2} {\bibfield
   {journal} {\bibinfo  {journal} {Adv. Theor. Math. Phys.}\ }\textbf {\bibinfo
  {volume} {2}},\ \bibinfo {pages} {253} (\bibinfo {year} {1998})},\ \Eprint
  {https://arxiv.org/abs/hep-th/9802150} {arXiv:hep-th/9802150} \BibitemShut
  {NoStop}%
\bibitem [{\citenamefont {Maldacena}(1999)}]{Maldacena1999}%
  \BibitemOpen
  \bibfield  {author} {\bibinfo {author} {\bibfnamefont {J.}~\bibnamefont
  {Maldacena}},\ }\href {https://doi.org/10.1023/a:1026654312961} {\bibfield
  {journal} {\bibinfo  {journal} {International Journal of Theoretical
  Physics}\ }\textbf {\bibinfo {volume} {38}},\ \bibinfo {pages} {1113–1133}
  (\bibinfo {year} {1999})}\BibitemShut {NoStop}%
\bibitem [{\citenamefont {Hayden}\ and\ \citenamefont
  {Preskill}(2007)}]{HaydenPreskill2007}%
  \BibitemOpen
  \bibfield  {author} {\bibinfo {author} {\bibfnamefont {P.}~\bibnamefont
  {Hayden}}\ and\ \bibinfo {author} {\bibfnamefont {J.}~\bibnamefont
  {Preskill}},\ }\bibfield  {title} {\bibinfo {title} {Black holes as mirrors:
  quantum information in random subsystems},\ }\href
  {https://doi.org/10.1088/1126-6708/2007/09/120} {\bibfield  {journal}
  {\bibinfo  {journal} {Journal of High Energy Physics}\ }\textbf {\bibinfo
  {volume} {2007}},\ \bibinfo {pages} {120–120} (\bibinfo {year}
  {2007})}\BibitemShut {NoStop}%
\bibitem [{\citenamefont {Sekino}\ and\ \citenamefont
  {Susskind}(2008)}]{SekinoSussking2008}%
  \BibitemOpen
  \bibfield  {author} {\bibinfo {author} {\bibfnamefont {Y.}~\bibnamefont
  {Sekino}}\ and\ \bibinfo {author} {\bibfnamefont {L.}~\bibnamefont
  {Susskind}},\ }\bibfield  {title} {\bibinfo {title} {Fast scramblers},\
  }\href {https://doi.org/10.1088/1126-6708/2008/10/065} {\bibfield  {journal}
  {\bibinfo  {journal} {Journal of High Energy Physics}\ }\textbf {\bibinfo
  {volume} {2008}},\ \bibinfo {pages} {065–065} (\bibinfo {year}
  {2008})}\BibitemShut {NoStop}%
\bibitem [{\citenamefont {Shenker}\ and\ \citenamefont
  {Stanford}(2014)}]{Shenker_2014}%
  \BibitemOpen
  \bibfield  {author} {\bibinfo {author} {\bibfnamefont {S.~H.}\ \bibnamefont
  {Shenker}}\ and\ \bibinfo {author} {\bibfnamefont {D.}~\bibnamefont
  {Stanford}},\ }\bibfield  {title} {\bibinfo {title} {Black holes and the
  butterfly effect},\ }\bibfield  {journal} {\bibinfo  {journal} {Journal of
  High Energy Physics}\ }\textbf {\bibinfo {volume} {2014}},\ \href
  {https://doi.org/10.1007/jhep03(2014)067} {10.1007/jhep03(2014)067} (\bibinfo
  {year} {2014})\BibitemShut {NoStop}%
\bibitem [{\citenamefont {Sachdev}\ and\ \citenamefont
  {Ye}(1993)}]{SachdevYe1993}%
  \BibitemOpen
  \bibfield  {author} {\bibinfo {author} {\bibfnamefont {S.}~\bibnamefont
  {Sachdev}}\ and\ \bibinfo {author} {\bibfnamefont {J.}~\bibnamefont {Ye}},\
  }\bibfield  {title} {\bibinfo {title} {Gapless spin-fluid ground state in a
  random quantum heisenberg magnet},\ }\href
  {https://doi.org/10.1103/physrevlett.70.3339} {\bibfield  {journal} {\bibinfo
   {journal} {Physical Review Letters}\ }\textbf {\bibinfo {volume} {70}},\
  \bibinfo {pages} {3339} (\bibinfo {year} {1993})}\BibitemShut {NoStop}%
\bibitem [{\citenamefont {Kitaev}(2015)}]{Kitaev}%
  \BibitemOpen
  \bibfield  {author} {\bibinfo {author} {\bibfnamefont {A.}~\bibnamefont
  {Kitaev}},\ }\bibfield  {title} {\bibinfo {title} {A simple model of quantum
  holography},\ }\href@noop {} {\bibfield  {journal} {\bibinfo  {journal}
  {Talks at KITP}\ } (\bibinfo {year} {2015})}\BibitemShut {NoStop}%
\bibitem [{\citenamefont {Lashkari}\ \emph {et~al.}(2013)\citenamefont
  {Lashkari}, \citenamefont {Stanford}, \citenamefont {Hastings}, \citenamefont
  {Osborne},\ and\ \citenamefont {Hayden}}]{LashkariHayden2013}%
  \BibitemOpen
  \bibfield  {author} {\bibinfo {author} {\bibfnamefont {N.}~\bibnamefont
  {Lashkari}}, \bibinfo {author} {\bibfnamefont {D.}~\bibnamefont {Stanford}},
  \bibinfo {author} {\bibfnamefont {M.}~\bibnamefont {Hastings}}, \bibinfo
  {author} {\bibfnamefont {T.}~\bibnamefont {Osborne}},\ and\ \bibinfo {author}
  {\bibfnamefont {P.}~\bibnamefont {Hayden}},\ }\bibfield  {title} {\bibinfo
  {title} {Towards the fast scrambling conjecture},\ }\bibfield  {journal}
  {\bibinfo  {journal} {Journal of High Energy Physics}\ }\textbf {\bibinfo
  {volume} {2013}},\ \href {https://doi.org/10.1007/jhep04(2013)022}
  {10.1007/jhep04(2013)022} (\bibinfo {year} {2013})\BibitemShut {NoStop}%
\bibitem [{\citenamefont {Roberts}\ and\ \citenamefont
  {Stanford}(2015)}]{RobertsStanford2015}%
  \BibitemOpen
  \bibfield  {author} {\bibinfo {author} {\bibfnamefont {D.~A.}\ \bibnamefont
  {Roberts}}\ and\ \bibinfo {author} {\bibfnamefont {D.}~\bibnamefont
  {Stanford}},\ }\bibfield  {title} {\bibinfo {title} {Diagnosing chaos using
  four-point functions in two-dimensional conformal field theory},\ }\href
  {https://doi.org/10.1103/PhysRevLett.115.131603} {\bibfield  {journal}
  {\bibinfo  {journal} {Phys. Rev. Lett.}\ }\textbf {\bibinfo {volume} {115}},\
  \bibinfo {pages} {131603} (\bibinfo {year} {2015})}\BibitemShut {NoStop}%
\bibitem [{\citenamefont {Cotler}\ \emph {et~al.}(2017)\citenamefont {Cotler},
  \citenamefont {Gur-Ari}, \citenamefont {Hanada}, \citenamefont {Polchinski},
  \citenamefont {Saad}, \citenamefont {Shenker}, \citenamefont {Stanford},
  \citenamefont {Streicher},\ and\ \citenamefont {Tezuka}}]{CotlerTezuka2017}%
  \BibitemOpen
  \bibfield  {author} {\bibinfo {author} {\bibfnamefont {J.~S.}\ \bibnamefont
  {Cotler}}, \bibinfo {author} {\bibfnamefont {G.}~\bibnamefont {Gur-Ari}},
  \bibinfo {author} {\bibfnamefont {M.}~\bibnamefont {Hanada}}, \bibinfo
  {author} {\bibfnamefont {J.}~\bibnamefont {Polchinski}}, \bibinfo {author}
  {\bibfnamefont {P.}~\bibnamefont {Saad}}, \bibinfo {author} {\bibfnamefont
  {S.~H.}\ \bibnamefont {Shenker}}, \bibinfo {author} {\bibfnamefont
  {D.}~\bibnamefont {Stanford}}, \bibinfo {author} {\bibfnamefont
  {A.}~\bibnamefont {Streicher}},\ and\ \bibinfo {author} {\bibfnamefont
  {M.}~\bibnamefont {Tezuka}},\ }\bibfield  {title} {\bibinfo {title} {Black
  holes and random matrices},\ }\bibfield  {journal} {\bibinfo  {journal}
  {Journal of High Energy Physics}\ }\textbf {\bibinfo {volume} {2017}},\ \href
  {https://doi.org/10.1007/jhep05(2017)118} {10.1007/jhep05(2017)118} (\bibinfo
  {year} {2017})\BibitemShut {NoStop}%
\bibitem [{\citenamefont {Roberts}\ \emph {et~al.}(2015)\citenamefont
  {Roberts}, \citenamefont {Stanford},\ and\ \citenamefont
  {Susskind}}]{RobertsSussking2015}%
  \BibitemOpen
  \bibfield  {author} {\bibinfo {author} {\bibfnamefont {D.~A.}\ \bibnamefont
  {Roberts}}, \bibinfo {author} {\bibfnamefont {D.}~\bibnamefont {Stanford}},\
  and\ \bibinfo {author} {\bibfnamefont {L.}~\bibnamefont {Susskind}},\
  }\bibfield  {title} {\bibinfo {title} {Localized shocks},\ }\bibfield
  {journal} {\bibinfo  {journal} {Journal of High Energy Physics}\ }\textbf
  {\bibinfo {volume} {2015}},\ \href {https://doi.org/10.1007/jhep03(2015)051}
  {10.1007/jhep03(2015)051} (\bibinfo {year} {2015})\BibitemShut {NoStop}%
\bibitem [{\citenamefont {Hosur}\ \emph {et~al.}(2016)\citenamefont {Hosur},
  \citenamefont {Qi}, \citenamefont {Roberts},\ and\ \citenamefont
  {Yoshida}}]{HosurYoshida2016}%
  \BibitemOpen
  \bibfield  {author} {\bibinfo {author} {\bibfnamefont {P.}~\bibnamefont
  {Hosur}}, \bibinfo {author} {\bibfnamefont {X.-L.}\ \bibnamefont {Qi}},
  \bibinfo {author} {\bibfnamefont {D.~A.}\ \bibnamefont {Roberts}},\ and\
  \bibinfo {author} {\bibfnamefont {B.}~\bibnamefont {Yoshida}},\ }\bibfield
  {title} {\bibinfo {title} {Chaos in quantum channels},\ }\bibfield  {journal}
  {\bibinfo  {journal} {Journal of High Energy Physics}\ }\textbf {\bibinfo
  {volume} {2016}},\ \href {https://doi.org/10.1007/jhep02(2016)004}
  {10.1007/jhep02(2016)004} (\bibinfo {year} {2016})\BibitemShut {NoStop}%
\bibitem [{\citenamefont {Borgonovi}\ \emph {et~al.}(2019)\citenamefont
  {Borgonovi}, \citenamefont {Izrailev},\ and\ \citenamefont
  {Santos}}]{Borgonovi2019}%
  \BibitemOpen
  \bibfield  {author} {\bibinfo {author} {\bibfnamefont {F.}~\bibnamefont
  {Borgonovi}}, \bibinfo {author} {\bibfnamefont {F.~M.}\ \bibnamefont
  {Izrailev}},\ and\ \bibinfo {author} {\bibfnamefont {L.~F.}\ \bibnamefont
  {Santos}},\ }\bibfield  {title} {\bibinfo {title} {Timescales in the quench
  dynamics of many-body quantum systems: Participation ratio versus out-of-time
  ordered correlator},\ }\href {https://doi.org/10.1103/PhysRevE.99.052143}
  {\bibfield  {journal} {\bibinfo  {journal} {Phys. Rev. E}\ }\textbf {\bibinfo
  {volume} {99}},\ \bibinfo {pages} {052143} (\bibinfo {year}
  {2019})}\BibitemShut {NoStop}%
\bibitem [{\citenamefont {Li}\ \emph {et~al.}(2017)\citenamefont {Li},
  \citenamefont {Fan}, \citenamefont {Wang}, \citenamefont {Ye}, \citenamefont
  {Zeng}, \citenamefont {Zhai}, \citenamefont {Peng},\ and\ \citenamefont
  {Du}}]{LiDu2017}%
  \BibitemOpen
  \bibfield  {author} {\bibinfo {author} {\bibfnamefont {J.}~\bibnamefont
  {Li}}, \bibinfo {author} {\bibfnamefont {R.}~\bibnamefont {Fan}}, \bibinfo
  {author} {\bibfnamefont {H.}~\bibnamefont {Wang}}, \bibinfo {author}
  {\bibfnamefont {B.}~\bibnamefont {Ye}}, \bibinfo {author} {\bibfnamefont
  {B.}~\bibnamefont {Zeng}}, \bibinfo {author} {\bibfnamefont {H.}~\bibnamefont
  {Zhai}}, \bibinfo {author} {\bibfnamefont {X.}~\bibnamefont {Peng}},\ and\
  \bibinfo {author} {\bibfnamefont {J.}~\bibnamefont {Du}},\ }\bibfield
  {title} {\bibinfo {title} {Measuring out-of-time-order correlators on a
  nuclear magnetic resonance quantum simulator},\ }\href
  {https://doi.org/10.1103/PhysRevX.7.031011} {\bibfield  {journal} {\bibinfo
  {journal} {Phys. Rev. X}\ }\textbf {\bibinfo {volume} {7}},\ \bibinfo {pages}
  {031011} (\bibinfo {year} {2017})}\BibitemShut {NoStop}%
\bibitem [{\citenamefont {Gärttner}\ \emph {et~al.}(2017)\citenamefont
  {Gärttner}, \citenamefont {Bohnet}, \citenamefont {Safavi-Naini},
  \citenamefont {Wall}, \citenamefont {Bollinger},\ and\ \citenamefont
  {Rey}}]{GarttnerRey2017}%
  \BibitemOpen
  \bibfield  {author} {\bibinfo {author} {\bibfnamefont {M.}~\bibnamefont
  {Gärttner}}, \bibinfo {author} {\bibfnamefont {J.~G.}\ \bibnamefont
  {Bohnet}}, \bibinfo {author} {\bibfnamefont {A.}~\bibnamefont
  {Safavi-Naini}}, \bibinfo {author} {\bibfnamefont {M.~L.}\ \bibnamefont
  {Wall}}, \bibinfo {author} {\bibfnamefont {J.~J.}\ \bibnamefont
  {Bollinger}},\ and\ \bibinfo {author} {\bibfnamefont {A.~M.}\ \bibnamefont
  {Rey}},\ }\bibfield  {title} {\bibinfo {title} {Measuring out-of-time-order
  correlations and multiple quantum spectra in a trapped-ion quantum magnet},\
  }\href {https://doi.org/10.1038/nphys4119} {\bibfield  {journal} {\bibinfo
  {journal} {Nature Physics}\ }\textbf {\bibinfo {volume} {13}},\ \bibinfo
  {pages} {781–786} (\bibinfo {year} {2017})}\BibitemShut {NoStop}%
\bibitem [{\citenamefont {Landsman}\ \emph {et~al.}(2019)\citenamefont
  {Landsman}, \citenamefont {Figgatt}, \citenamefont {Schuster}, \citenamefont
  {Linke}, \citenamefont {Yoshida}, \citenamefont {Yao},\ and\ \citenamefont
  {Monroe}}]{LandsmanMonroe2019}%
  \BibitemOpen
  \bibfield  {author} {\bibinfo {author} {\bibfnamefont {K.~A.}\ \bibnamefont
  {Landsman}}, \bibinfo {author} {\bibfnamefont {C.}~\bibnamefont {Figgatt}},
  \bibinfo {author} {\bibfnamefont {T.}~\bibnamefont {Schuster}}, \bibinfo
  {author} {\bibfnamefont {N.~M.}\ \bibnamefont {Linke}}, \bibinfo {author}
  {\bibfnamefont {B.}~\bibnamefont {Yoshida}}, \bibinfo {author} {\bibfnamefont
  {N.~Y.}\ \bibnamefont {Yao}},\ and\ \bibinfo {author} {\bibfnamefont
  {C.}~\bibnamefont {Monroe}},\ }\bibfield  {title} {\bibinfo {title} {Verified
  quantum information scrambling},\ }\href
  {https://doi.org/10.1038/s41586-019-0952-6} {\bibfield  {journal} {\bibinfo
  {journal} {Nature}\ }\textbf {\bibinfo {volume} {567}},\ \bibinfo {pages}
  {61–65} (\bibinfo {year} {2019})}\BibitemShut {NoStop}%
\bibitem [{\citenamefont {Niknam}\ \emph {et~al.}(2020)\citenamefont {Niknam},
  \citenamefont {Santos},\ and\ \citenamefont {Cory}}]{Niknam2020}%
  \BibitemOpen
  \bibfield  {author} {\bibinfo {author} {\bibfnamefont {M.}~\bibnamefont
  {Niknam}}, \bibinfo {author} {\bibfnamefont {L.~F.}\ \bibnamefont {Santos}},\
  and\ \bibinfo {author} {\bibfnamefont {D.~G.}\ \bibnamefont {Cory}},\
  }\bibfield  {title} {\bibinfo {title} {Sensitivity of quantum information to
  environment perturbations measured with a nonlocal out-of-time-order
  correlation function},\ }\href
  {https://doi.org/10.1103/PhysRevResearch.2.013200} {\bibfield  {journal}
  {\bibinfo  {journal} {Phys. Rev. Research}\ }\textbf {\bibinfo {volume}
  {2}},\ \bibinfo {pages} {013200} (\bibinfo {year} {2020})}\BibitemShut
  {NoStop}%
\bibitem [{\citenamefont {Joshi}\ \emph {et~al.}(2020)\citenamefont {Joshi},
  \citenamefont {Elben}, \citenamefont {Vermersch}, \citenamefont {Brydges},
  \citenamefont {Maier}, \citenamefont {Zoller}, \citenamefont {Blatt},\ and\
  \citenamefont {Roos}}]{JoshiRoos2020}%
  \BibitemOpen
  \bibfield  {author} {\bibinfo {author} {\bibfnamefont {M.~K.}\ \bibnamefont
  {Joshi}}, \bibinfo {author} {\bibfnamefont {A.}~\bibnamefont {Elben}},
  \bibinfo {author} {\bibfnamefont {B.}~\bibnamefont {Vermersch}}, \bibinfo
  {author} {\bibfnamefont {T.}~\bibnamefont {Brydges}}, \bibinfo {author}
  {\bibfnamefont {C.}~\bibnamefont {Maier}}, \bibinfo {author} {\bibfnamefont
  {P.}~\bibnamefont {Zoller}}, \bibinfo {author} {\bibfnamefont
  {R.}~\bibnamefont {Blatt}},\ and\ \bibinfo {author} {\bibfnamefont {C.~F.}\
  \bibnamefont {Roos}},\ }\bibfield  {title} {\bibinfo {title} {Quantum
  information scrambling in a trapped-ion quantum simulator with tunable range
  interactions},\ }\href {https://doi.org/10.1103/PhysRevLett.124.240505}
  {\bibfield  {journal} {\bibinfo  {journal} {Phys. Rev. Lett.}\ }\textbf
  {\bibinfo {volume} {124}},\ \bibinfo {pages} {240505} (\bibinfo {year}
  {2020})}\BibitemShut {NoStop}%
\bibitem [{\citenamefont {Blok}\ \emph {et~al.}(2021)\citenamefont {Blok},
  \citenamefont {Ramasesh}, \citenamefont {Schuster}, \citenamefont {O'Brien},
  \citenamefont {Kreikebaum}, \citenamefont {Dahlen}, \citenamefont {Morvan},
  \citenamefont {Yoshida}, \citenamefont {Yao},\ and\ \citenamefont
  {Siddiqi}}]{BlokSiddiqi2021}%
  \BibitemOpen
  \bibfield  {author} {\bibinfo {author} {\bibfnamefont {M.~S.}\ \bibnamefont
  {Blok}}, \bibinfo {author} {\bibfnamefont {V.~V.}\ \bibnamefont {Ramasesh}},
  \bibinfo {author} {\bibfnamefont {T.}~\bibnamefont {Schuster}}, \bibinfo
  {author} {\bibfnamefont {K.}~\bibnamefont {O'Brien}}, \bibinfo {author}
  {\bibfnamefont {J.~M.}\ \bibnamefont {Kreikebaum}}, \bibinfo {author}
  {\bibfnamefont {D.}~\bibnamefont {Dahlen}}, \bibinfo {author} {\bibfnamefont
  {A.}~\bibnamefont {Morvan}}, \bibinfo {author} {\bibfnamefont
  {B.}~\bibnamefont {Yoshida}}, \bibinfo {author} {\bibfnamefont {N.~Y.}\
  \bibnamefont {Yao}},\ and\ \bibinfo {author} {\bibfnamefont {I.}~\bibnamefont
  {Siddiqi}},\ }\bibfield  {title} {\bibinfo {title} {Quantum information
  scrambling on a superconducting qutrit processor},\ }\href
  {https://doi.org/10.1103/PhysRevX.11.021010} {\bibfield  {journal} {\bibinfo
  {journal} {Phys. Rev. X}\ }\textbf {\bibinfo {volume} {11}},\ \bibinfo
  {pages} {021010} (\bibinfo {year} {2021})}\BibitemShut {NoStop}%
\bibitem [{\citenamefont {Mi}\ \emph {et~al.}(2021)\citenamefont {Mi},
  \citenamefont {Roushan}, \citenamefont {Quintana}, \citenamefont {Mandra},
  \citenamefont {Marshall}, \citenamefont {Neill}, \citenamefont {Arute},
  \citenamefont {Arya}, \citenamefont {Atalaya}, \citenamefont {Babbush},
  \citenamefont {Bardin}, \citenamefont {Barends}, \citenamefont {Bengtsson},
  \citenamefont {Boixo}, \citenamefont {Bourassa}, \citenamefont {Broughton},
  \citenamefont {Buckley}, \citenamefont {Buell}, \citenamefont {Burkett},
  \citenamefont {Bushnell}, \citenamefont {Chen}, \citenamefont {Chiaro},
  \citenamefont {Collins}, \citenamefont {Courtney}, \citenamefont {Demura},
  \citenamefont {Derk}, \citenamefont {Dunsworth}, \citenamefont {Eppens},
  \citenamefont {Erickson}, \citenamefont {Farhi}, \citenamefont {Fowler},
  \citenamefont {Foxen}, \citenamefont {Gidney}, \citenamefont {Giustina},
  \citenamefont {Gross}, \citenamefont {Harrigan}, \citenamefont {Harrington},
  \citenamefont {Hilton}, \citenamefont {Ho}, \citenamefont {Hong},
  \citenamefont {Huang}, \citenamefont {Huggins}, \citenamefont {Ioffe},
  \citenamefont {Isakov}, \citenamefont {Jeffrey}, \citenamefont {Jiang},
  \citenamefont {Jones}, \citenamefont {Kafri}, \citenamefont {Kelly},
  \citenamefont {Kim}, \citenamefont {Kitaev}, \citenamefont {Klimov},
  \citenamefont {Korotkov}, \citenamefont {Kostritsa}, \citenamefont
  {Landhuis}, \citenamefont {Laptev}, \citenamefont {Lucero}, \citenamefont
  {Martin}, \citenamefont {McClean}, \citenamefont {McCourt}, \citenamefont
  {McEwen}, \citenamefont {Megrant}, \citenamefont {Miao}, \citenamefont
  {Mohseni}, \citenamefont {Mruczkiewicz}, \citenamefont {Mutus}, \citenamefont
  {Naaman}, \citenamefont {Neeley}, \citenamefont {Newman}, \citenamefont
  {Niu}, \citenamefont {O'Brien}, \citenamefont {Opremcak}, \citenamefont
  {Ostby}, \citenamefont {Pato}, \citenamefont {Petukhov}, \citenamefont
  {Redd}, \citenamefont {Rubin}, \citenamefont {Sank}, \citenamefont
  {Satzinger}, \citenamefont {Shvarts}, \citenamefont {Strain}, \citenamefont
  {Szalay}, \citenamefont {Trevithick}, \citenamefont {Villalonga},
  \citenamefont {White}, \citenamefont {Yao}, \citenamefont {Yeh},
  \citenamefont {Zalcman}, \citenamefont {Neven}, \citenamefont {Aleiner},
  \citenamefont {Kechedzhi}, \citenamefont {Smelyanskiy},\ and\ \citenamefont
  {Chen}}]{MiYu2021}%
  \BibitemOpen
  \bibfield  {author} {\bibinfo {author} {\bibfnamefont {X.}~\bibnamefont
  {Mi}}, \bibinfo {author} {\bibfnamefont {P.}~\bibnamefont {Roushan}},
  \bibinfo {author} {\bibfnamefont {C.}~\bibnamefont {Quintana}}, \bibinfo
  {author} {\bibfnamefont {S.}~\bibnamefont {Mandra}}, \bibinfo {author}
  {\bibfnamefont {J.}~\bibnamefont {Marshall}}, \bibinfo {author}
  {\bibfnamefont {C.}~\bibnamefont {Neill}}, \bibinfo {author} {\bibfnamefont
  {F.}~\bibnamefont {Arute}}, \bibinfo {author} {\bibfnamefont
  {K.}~\bibnamefont {Arya}}, \bibinfo {author} {\bibfnamefont {J.}~\bibnamefont
  {Atalaya}}, \bibinfo {author} {\bibfnamefont {R.}~\bibnamefont {Babbush}},
  \bibinfo {author} {\bibfnamefont {J.~C.}\ \bibnamefont {Bardin}}, \bibinfo
  {author} {\bibfnamefont {R.}~\bibnamefont {Barends}}, \bibinfo {author}
  {\bibfnamefont {A.}~\bibnamefont {Bengtsson}}, \bibinfo {author}
  {\bibfnamefont {S.}~\bibnamefont {Boixo}}, \bibinfo {author} {\bibfnamefont
  {A.}~\bibnamefont {Bourassa}}, \bibinfo {author} {\bibfnamefont
  {M.}~\bibnamefont {Broughton}}, \bibinfo {author} {\bibfnamefont {B.~B.}\
  \bibnamefont {Buckley}}, \bibinfo {author} {\bibfnamefont {D.~A.}\
  \bibnamefont {Buell}}, \bibinfo {author} {\bibfnamefont {B.}~\bibnamefont
  {Burkett}}, \bibinfo {author} {\bibfnamefont {N.}~\bibnamefont {Bushnell}},
  \bibinfo {author} {\bibfnamefont {Z.}~\bibnamefont {Chen}}, \bibinfo {author}
  {\bibfnamefont {B.}~\bibnamefont {Chiaro}}, \bibinfo {author} {\bibfnamefont
  {R.}~\bibnamefont {Collins}}, \bibinfo {author} {\bibfnamefont
  {W.}~\bibnamefont {Courtney}}, \bibinfo {author} {\bibfnamefont
  {S.}~\bibnamefont {Demura}}, \bibinfo {author} {\bibfnamefont {A.~R.}\
  \bibnamefont {Derk}}, \bibinfo {author} {\bibfnamefont {A.}~\bibnamefont
  {Dunsworth}}, \bibinfo {author} {\bibfnamefont {D.}~\bibnamefont {Eppens}},
  \bibinfo {author} {\bibfnamefont {C.}~\bibnamefont {Erickson}}, \bibinfo
  {author} {\bibfnamefont {E.}~\bibnamefont {Farhi}}, \bibinfo {author}
  {\bibfnamefont {A.~G.}\ \bibnamefont {Fowler}}, \bibinfo {author}
  {\bibfnamefont {B.}~\bibnamefont {Foxen}}, \bibinfo {author} {\bibfnamefont
  {C.}~\bibnamefont {Gidney}}, \bibinfo {author} {\bibfnamefont
  {M.}~\bibnamefont {Giustina}}, \bibinfo {author} {\bibfnamefont {J.~A.}\
  \bibnamefont {Gross}}, \bibinfo {author} {\bibfnamefont {M.~P.}\ \bibnamefont
  {Harrigan}}, \bibinfo {author} {\bibfnamefont {S.~D.}\ \bibnamefont
  {Harrington}}, \bibinfo {author} {\bibfnamefont {J.}~\bibnamefont {Hilton}},
  \bibinfo {author} {\bibfnamefont {A.}~\bibnamefont {Ho}}, \bibinfo {author}
  {\bibfnamefont {S.}~\bibnamefont {Hong}}, \bibinfo {author} {\bibfnamefont
  {T.}~\bibnamefont {Huang}}, \bibinfo {author} {\bibfnamefont {W.~J.}\
  \bibnamefont {Huggins}}, \bibinfo {author} {\bibfnamefont {L.~B.}\
  \bibnamefont {Ioffe}}, \bibinfo {author} {\bibfnamefont {S.~V.}\ \bibnamefont
  {Isakov}}, \bibinfo {author} {\bibfnamefont {E.}~\bibnamefont {Jeffrey}},
  \bibinfo {author} {\bibfnamefont {Z.}~\bibnamefont {Jiang}}, \bibinfo
  {author} {\bibfnamefont {C.}~\bibnamefont {Jones}}, \bibinfo {author}
  {\bibfnamefont {D.}~\bibnamefont {Kafri}}, \bibinfo {author} {\bibfnamefont
  {J.}~\bibnamefont {Kelly}}, \bibinfo {author} {\bibfnamefont
  {S.}~\bibnamefont {Kim}}, \bibinfo {author} {\bibfnamefont {A.}~\bibnamefont
  {Kitaev}}, \bibinfo {author} {\bibfnamefont {P.~V.}\ \bibnamefont {Klimov}},
  \bibinfo {author} {\bibfnamefont {A.~N.}\ \bibnamefont {Korotkov}}, \bibinfo
  {author} {\bibfnamefont {F.}~\bibnamefont {Kostritsa}}, \bibinfo {author}
  {\bibfnamefont {D.}~\bibnamefont {Landhuis}}, \bibinfo {author}
  {\bibfnamefont {P.}~\bibnamefont {Laptev}}, \bibinfo {author} {\bibfnamefont
  {E.}~\bibnamefont {Lucero}}, \bibinfo {author} {\bibfnamefont
  {O.}~\bibnamefont {Martin}}, \bibinfo {author} {\bibfnamefont {J.~R.}\
  \bibnamefont {McClean}}, \bibinfo {author} {\bibfnamefont {T.}~\bibnamefont
  {McCourt}}, \bibinfo {author} {\bibfnamefont {M.}~\bibnamefont {McEwen}},
  \bibinfo {author} {\bibfnamefont {A.}~\bibnamefont {Megrant}}, \bibinfo
  {author} {\bibfnamefont {K.~C.}\ \bibnamefont {Miao}}, \bibinfo {author}
  {\bibfnamefont {M.}~\bibnamefont {Mohseni}}, \bibinfo {author} {\bibfnamefont
  {W.}~\bibnamefont {Mruczkiewicz}}, \bibinfo {author} {\bibfnamefont
  {J.}~\bibnamefont {Mutus}}, \bibinfo {author} {\bibfnamefont
  {O.}~\bibnamefont {Naaman}}, \bibinfo {author} {\bibfnamefont
  {M.}~\bibnamefont {Neeley}}, \bibinfo {author} {\bibfnamefont
  {M.}~\bibnamefont {Newman}}, \bibinfo {author} {\bibfnamefont {M.~Y.}\
  \bibnamefont {Niu}}, \bibinfo {author} {\bibfnamefont {T.~E.}\ \bibnamefont
  {O'Brien}}, \bibinfo {author} {\bibfnamefont {A.}~\bibnamefont {Opremcak}},
  \bibinfo {author} {\bibfnamefont {E.}~\bibnamefont {Ostby}}, \bibinfo
  {author} {\bibfnamefont {B.}~\bibnamefont {Pato}}, \bibinfo {author}
  {\bibfnamefont {A.}~\bibnamefont {Petukhov}}, \bibinfo {author}
  {\bibfnamefont {N.}~\bibnamefont {Redd}}, \bibinfo {author} {\bibfnamefont
  {N.~C.}\ \bibnamefont {Rubin}}, \bibinfo {author} {\bibfnamefont
  {D.}~\bibnamefont {Sank}}, \bibinfo {author} {\bibfnamefont {K.~J.}\
  \bibnamefont {Satzinger}}, \bibinfo {author} {\bibfnamefont {V.}~\bibnamefont
  {Shvarts}}, \bibinfo {author} {\bibfnamefont {D.}~\bibnamefont {Strain}},
  \bibinfo {author} {\bibfnamefont {M.}~\bibnamefont {Szalay}}, \bibinfo
  {author} {\bibfnamefont {M.~D.}\ \bibnamefont {Trevithick}}, \bibinfo
  {author} {\bibfnamefont {B.}~\bibnamefont {Villalonga}}, \bibinfo {author}
  {\bibfnamefont {T.}~\bibnamefont {White}}, \bibinfo {author} {\bibfnamefont
  {Z.~J.}\ \bibnamefont {Yao}}, \bibinfo {author} {\bibfnamefont
  {P.}~\bibnamefont {Yeh}}, \bibinfo {author} {\bibfnamefont {A.}~\bibnamefont
  {Zalcman}}, \bibinfo {author} {\bibfnamefont {H.}~\bibnamefont {Neven}},
  \bibinfo {author} {\bibfnamefont {I.}~\bibnamefont {Aleiner}}, \bibinfo
  {author} {\bibfnamefont {K.}~\bibnamefont {Kechedzhi}}, \bibinfo {author}
  {\bibfnamefont {V.}~\bibnamefont {Smelyanskiy}},\ and\ \bibinfo {author}
  {\bibfnamefont {Y.}~\bibnamefont {Chen}},\ }\href@noop {} {\bibinfo {title}
  {Information scrambling in computationally complex quantum circuits}}
  (\bibinfo {year} {2021}),\ \Eprint {https://arxiv.org/abs/2101.08870}
  {arXiv:2101.08870 [quant-ph]} \BibitemShut {NoStop}%
\bibitem [{\citenamefont {Braumüller}\ \emph {et~al.}(2021)\citenamefont
  {Braumüller}, \citenamefont {Karamlou}, \citenamefont {Yanay}, \citenamefont
  {Kannan}, \citenamefont {Kim}, \citenamefont {Kjaergaard}, \citenamefont
  {Melville}, \citenamefont {Niedzielski}, \citenamefont {Sung}, \citenamefont
  {Vepsäläinen}, \citenamefont {Winik}, \citenamefont {Yoder}, \citenamefont
  {Orlando}, \citenamefont {Gustavsson}, \citenamefont {Tahan},\ and\
  \citenamefont {Oliver}}]{Jochen2021}%
  \BibitemOpen
  \bibfield  {author} {\bibinfo {author} {\bibfnamefont {J.}~\bibnamefont
  {Braumüller}}, \bibinfo {author} {\bibfnamefont {A.~H.}\ \bibnamefont
  {Karamlou}}, \bibinfo {author} {\bibfnamefont {Y.}~\bibnamefont {Yanay}},
  \bibinfo {author} {\bibfnamefont {B.}~\bibnamefont {Kannan}}, \bibinfo
  {author} {\bibfnamefont {D.}~\bibnamefont {Kim}}, \bibinfo {author}
  {\bibfnamefont {M.}~\bibnamefont {Kjaergaard}}, \bibinfo {author}
  {\bibfnamefont {A.}~\bibnamefont {Melville}}, \bibinfo {author}
  {\bibfnamefont {B.~M.}\ \bibnamefont {Niedzielski}}, \bibinfo {author}
  {\bibfnamefont {Y.}~\bibnamefont {Sung}}, \bibinfo {author} {\bibfnamefont
  {A.}~\bibnamefont {Vepsäläinen}}, \bibinfo {author} {\bibfnamefont
  {R.}~\bibnamefont {Winik}}, \bibinfo {author} {\bibfnamefont {J.~L.}\
  \bibnamefont {Yoder}}, \bibinfo {author} {\bibfnamefont {T.~P.}\ \bibnamefont
  {Orlando}}, \bibinfo {author} {\bibfnamefont {S.}~\bibnamefont {Gustavsson}},
  \bibinfo {author} {\bibfnamefont {C.}~\bibnamefont {Tahan}},\ and\ \bibinfo
  {author} {\bibfnamefont {W.~D.}\ \bibnamefont {Oliver}},\ }\href@noop {}
  {\bibinfo {title} {Probing quantum information propagation with
  out-of-time-ordered correlators}} (\bibinfo {year} {2021}),\ \Eprint
  {https://arxiv.org/abs/2102.11751} {arXiv:2102.11751 [quant-ph]} \BibitemShut
  {NoStop}%
\bibitem [{\citenamefont {Rozenbaum}\ \emph {et~al.}(2017)\citenamefont
  {Rozenbaum}, \citenamefont {Ganeshan},\ and\ \citenamefont
  {Galitski}}]{Galitski2017}%
  \BibitemOpen
  \bibfield  {author} {\bibinfo {author} {\bibfnamefont {E.~B.}\ \bibnamefont
  {Rozenbaum}}, \bibinfo {author} {\bibfnamefont {S.}~\bibnamefont
  {Ganeshan}},\ and\ \bibinfo {author} {\bibfnamefont {V.}~\bibnamefont
  {Galitski}},\ }\bibfield  {title} {\bibinfo {title} {Lyapunov exponent and
  out-of-time-ordered correlator's growth rate in a chaotic system},\ }\href
  {https://doi.org/10.1103/PhysRevLett.118.086801} {\bibfield  {journal}
  {\bibinfo  {journal} {Phys. Rev. Lett.}\ }\textbf {\bibinfo {volume} {118}},\
  \bibinfo {pages} {086801} (\bibinfo {year} {2017})}\BibitemShut {NoStop}%
\bibitem [{\citenamefont {Hashimoto}\ \emph {et~al.}(2017)\citenamefont
  {Hashimoto}, \citenamefont {Murata},\ and\ \citenamefont
  {Yoshii}}]{Hashimoto_2017}%
  \BibitemOpen
  \bibfield  {author} {\bibinfo {author} {\bibfnamefont {K.}~\bibnamefont
  {Hashimoto}}, \bibinfo {author} {\bibfnamefont {K.}~\bibnamefont {Murata}},\
  and\ \bibinfo {author} {\bibfnamefont {R.}~\bibnamefont {Yoshii}},\
  }\bibfield  {title} {\bibinfo {title} {Out-of-time-order correlators in
  quantum mechanics},\ }\bibfield  {journal} {\bibinfo  {journal} {Journal of
  High Energy Physics}\ }\textbf {\bibinfo {volume} {2017}},\ \href
  {https://doi.org/10.1007/jhep10(2017)138} {10.1007/jhep10(2017)138} (\bibinfo
  {year} {2017})\BibitemShut {NoStop}%
\bibitem [{\citenamefont {Cotler}\ \emph {et~al.}(2018)\citenamefont {Cotler},
  \citenamefont {Ding},\ and\ \citenamefont {Penington}}]{Cotler_2018}%
  \BibitemOpen
  \bibfield  {author} {\bibinfo {author} {\bibfnamefont {J.~S.}\ \bibnamefont
  {Cotler}}, \bibinfo {author} {\bibfnamefont {D.}~\bibnamefont {Ding}},\ and\
  \bibinfo {author} {\bibfnamefont {G.~R.}\ \bibnamefont {Penington}},\
  }\bibfield  {title} {\bibinfo {title} {Out-of-time-order operators and the
  butterfly effect},\ }\href {https://doi.org/10.1016/j.aop.2018.07.020}
  {\bibfield  {journal} {\bibinfo  {journal} {Annals of Physics}\ }\textbf
  {\bibinfo {volume} {396}},\ \bibinfo {pages} {318–333} (\bibinfo {year}
  {2018})}\BibitemShut {NoStop}%
\bibitem [{\citenamefont {Garc\'{\i}a-Mata}\ \emph {et~al.}(2018)\citenamefont
  {Garc\'{\i}a-Mata}, \citenamefont {Saraceno}, \citenamefont {Jalabert},
  \citenamefont {Roncaglia},\ and\ \citenamefont {Wisniacki}}]{Ignacio2018}%
  \BibitemOpen
  \bibfield  {author} {\bibinfo {author} {\bibfnamefont {I.}~\bibnamefont
  {Garc\'{\i}a-Mata}}, \bibinfo {author} {\bibfnamefont {M.}~\bibnamefont
  {Saraceno}}, \bibinfo {author} {\bibfnamefont {R.~A.}\ \bibnamefont
  {Jalabert}}, \bibinfo {author} {\bibfnamefont {A.~J.}\ \bibnamefont
  {Roncaglia}},\ and\ \bibinfo {author} {\bibfnamefont {D.~A.}\ \bibnamefont
  {Wisniacki}},\ }\bibfield  {title} {\bibinfo {title} {Chaos signatures in the
  short and long time behavior of the out-of-time ordered correlator},\ }\href
  {https://doi.org/10.1103/PhysRevLett.121.210601} {\bibfield  {journal}
  {\bibinfo  {journal} {Phys. Rev. Lett.}\ }\textbf {\bibinfo {volume} {121}},\
  \bibinfo {pages} {210601} (\bibinfo {year} {2018})}\BibitemShut {NoStop}%
\bibitem [{\citenamefont {Ch\'{a}vez-Carlos}\ \emph {et~al.}(2019)\citenamefont
  {Ch\'{a}vez-Carlos}, \citenamefont {L\'{o}pez-del Carpio}, \citenamefont
  {Bastarrachea-Magnani}, \citenamefont {Str\'{a}nsk\'{y}}, \citenamefont
  {Lerma-Hernández}, \citenamefont {Santos},\ and\ \citenamefont
  {Hirsch}}]{Ch_vez_Carlos_2019}%
  \BibitemOpen
  \bibfield  {author} {\bibinfo {author} {\bibfnamefont {J.}~\bibnamefont
  {Ch\'{a}vez-Carlos}}, \bibinfo {author} {\bibfnamefont {B.}~\bibnamefont
  {L\'{o}pez-del Carpio}}, \bibinfo {author} {\bibfnamefont {M.~A.}\
  \bibnamefont {Bastarrachea-Magnani}}, \bibinfo {author} {\bibfnamefont
  {P.}~\bibnamefont {Str\'{a}nsk\'{y}}}, \bibinfo {author} {\bibfnamefont
  {S.}~\bibnamefont {Lerma-Hernández}}, \bibinfo {author} {\bibfnamefont
  {L.~F.}\ \bibnamefont {Santos}},\ and\ \bibinfo {author} {\bibfnamefont
  {J.~G.}\ \bibnamefont {Hirsch}},\ }\bibfield  {title} {\bibinfo {title}
  {Quantum and classical lyapunov exponents in atom-field interaction
  systems},\ }\href {https://doi.org/10.1103/physrevlett.122.024101} {\bibfield
   {journal} {\bibinfo  {journal} {Physical Review Letters}\ }\textbf {\bibinfo
  {volume} {122}},\ \bibinfo {pages} {024101} (\bibinfo {year}
  {2019})}\BibitemShut {NoStop}%
\bibitem [{\citenamefont {Fortes}\ \emph {et~al.}(2019)\citenamefont {Fortes},
  \citenamefont {Garc\'{\i}a-Mata}, \citenamefont {Jalabert},\ and\
  \citenamefont {Wisniacki}}]{Fortes2019}%
  \BibitemOpen
  \bibfield  {author} {\bibinfo {author} {\bibfnamefont {E.~M.}\ \bibnamefont
  {Fortes}}, \bibinfo {author} {\bibfnamefont {I.}~\bibnamefont
  {Garc\'{\i}a-Mata}}, \bibinfo {author} {\bibfnamefont {R.~A.}\ \bibnamefont
  {Jalabert}},\ and\ \bibinfo {author} {\bibfnamefont {D.~A.}\ \bibnamefont
  {Wisniacki}},\ }\bibfield  {title} {\bibinfo {title} {Gauging classical and
  quantum integrability through out-of-time-ordered correlators},\ }\href
  {https://doi.org/10.1103/PhysRevE.100.042201} {\bibfield  {journal} {\bibinfo
   {journal} {Phys. Rev. E}\ }\textbf {\bibinfo {volume} {100}},\ \bibinfo
  {pages} {042201} (\bibinfo {year} {2019})}\BibitemShut {NoStop}%
\bibitem [{\citenamefont {Rammensee}\ \emph {et~al.}(2018)\citenamefont
  {Rammensee}, \citenamefont {Urbina},\ and\ \citenamefont
  {Richter}}]{Rammensee2018}%
  \BibitemOpen
  \bibfield  {author} {\bibinfo {author} {\bibfnamefont {J.}~\bibnamefont
  {Rammensee}}, \bibinfo {author} {\bibfnamefont {J.~D.}\ \bibnamefont
  {Urbina}},\ and\ \bibinfo {author} {\bibfnamefont {K.}~\bibnamefont
  {Richter}},\ }\bibfield  {title} {\bibinfo {title} {Many-body quantum
  interference and the saturation of out-of-time-order correlators},\ }\href
  {https://doi.org/10.1103/PhysRevLett.121.124101} {\bibfield  {journal}
  {\bibinfo  {journal} {Phys. Rev. Lett.}\ }\textbf {\bibinfo {volume} {121}},\
  \bibinfo {pages} {124101} (\bibinfo {year} {2018})}\BibitemShut {NoStop}%
\bibitem [{\citenamefont {Prakash}\ and\ \citenamefont
  {Lakshminarayan}(2020)}]{Prakash2020}%
  \BibitemOpen
  \bibfield  {author} {\bibinfo {author} {\bibfnamefont {R.}~\bibnamefont
  {Prakash}}\ and\ \bibinfo {author} {\bibfnamefont {A.}~\bibnamefont
  {Lakshminarayan}},\ }\bibfield  {title} {\bibinfo {title} {Scrambling in
  strongly chaotic weakly coupled bipartite systems: Universality beyond the
  ehrenfest timescale},\ }\href {https://doi.org/10.1103/PhysRevB.101.121108}
  {\bibfield  {journal} {\bibinfo  {journal} {Physical Review B}\ }\textbf
  {\bibinfo {volume} {101}},\ \bibinfo {pages} {121108(R)} (\bibinfo {year}
  {2020})}\BibitemShut {NoStop}%
\bibitem [{\citenamefont {Bergamasco}\ \emph {et~al.}(2019)\citenamefont
  {Bergamasco}, \citenamefont {Carlo},\ and\ \citenamefont
  {Rivas}}]{Bergamasco2019}%
  \BibitemOpen
  \bibfield  {author} {\bibinfo {author} {\bibfnamefont {P.~D.}\ \bibnamefont
  {Bergamasco}}, \bibinfo {author} {\bibfnamefont {G.~G.}\ \bibnamefont
  {Carlo}},\ and\ \bibinfo {author} {\bibfnamefont {A.~M.~F.}\ \bibnamefont
  {Rivas}},\ }\bibfield  {title} {\bibinfo {title} {Out-of-time ordered
  correlators, complexity, and entropy in bipartite systems},\ }\href
  {https://doi.org/10.1103/PhysRevResearch.1.033044} {\bibfield  {journal}
  {\bibinfo  {journal} {Physical Review Research}\ }\textbf {\bibinfo {volume}
  {1}},\ \bibinfo {pages} {033044} (\bibinfo {year} {2019})}\BibitemShut
  {NoStop}%
\bibitem [{\citenamefont {Rozenbaum}\ \emph {et~al.}(2020)\citenamefont
  {Rozenbaum}, \citenamefont {Bunimovich},\ and\ \citenamefont
  {Galitski}}]{Rozenbaum2020}%
  \BibitemOpen
  \bibfield  {author} {\bibinfo {author} {\bibfnamefont {E.~B.}\ \bibnamefont
  {Rozenbaum}}, \bibinfo {author} {\bibfnamefont {L.~A.}\ \bibnamefont
  {Bunimovich}},\ and\ \bibinfo {author} {\bibfnamefont {V.}~\bibnamefont
  {Galitski}},\ }\bibfield  {title} {\bibinfo {title} {Early-time exponential
  instabilities in nonchaotic quantum systems},\ }\href
  {https://doi.org/10.1103/PhysRevLett.125.014101} {\bibfield  {journal}
  {\bibinfo  {journal} {Phys. Rev. Lett.}\ }\textbf {\bibinfo {volume} {125}},\
  \bibinfo {pages} {014101} (\bibinfo {year} {2020})}\BibitemShut {NoStop}%
\bibitem [{\citenamefont {Wang}\ \emph {et~al.}(2020)\citenamefont {Wang},
  \citenamefont {Benenti}, \citenamefont {Casati},\ and\ \citenamefont
  {Wang}}]{Wang2020}%
  \BibitemOpen
  \bibfield  {author} {\bibinfo {author} {\bibfnamefont {J.}~\bibnamefont
  {Wang}}, \bibinfo {author} {\bibfnamefont {G.}~\bibnamefont {Benenti}},
  \bibinfo {author} {\bibfnamefont {G.}~\bibnamefont {Casati}},\ and\ \bibinfo
  {author} {\bibfnamefont {W.-g.}\ \bibnamefont {Wang}},\ }\bibfield  {title}
  {\bibinfo {title} {Complexity of quantum motion and quantum-classical
  correspondence: A phase-space approach},\ }\href
  {https://doi.org/10.1103/PhysRevResearch.2.043178} {\bibfield  {journal}
  {\bibinfo  {journal} {Phys. Rev. Research}\ }\textbf {\bibinfo {volume}
  {2}},\ \bibinfo {pages} {043178} (\bibinfo {year} {2020})}\BibitemShut
  {NoStop}%
\bibitem [{\citenamefont {Wang}\ \emph {et~al.}(2021)\citenamefont {Wang},
  \citenamefont {Benenti}, \citenamefont {Casati},\ and\ \citenamefont
  {Wang}}]{wang2020quantum}%
  \BibitemOpen
  \bibfield  {author} {\bibinfo {author} {\bibfnamefont {J.}~\bibnamefont
  {Wang}}, \bibinfo {author} {\bibfnamefont {G.}~\bibnamefont {Benenti}},
  \bibinfo {author} {\bibfnamefont {G.}~\bibnamefont {Casati}},\ and\ \bibinfo
  {author} {\bibfnamefont {W.-g.}\ \bibnamefont {Wang}},\ }\bibfield  {title}
  {\bibinfo {title} {Quantum chaos and the correspondence principle},\ }\href
  {https://doi.org/10.1103/PhysRevE.103.L030201} {\bibfield  {journal}
  {\bibinfo  {journal} {Phys. Rev. E}\ }\textbf {\bibinfo {volume} {103}},\
  \bibinfo {pages} {L030201} (\bibinfo {year} {2021})}\BibitemShut {NoStop}%
\bibitem [{\citenamefont {Rakovszky}\ \emph {et~al.}(2018)\citenamefont
  {Rakovszky}, \citenamefont {Pollmann},\ and\ \citenamefont {von
  Keyserlingk}}]{RakovszkyKeyserlingk2018}%
  \BibitemOpen
  \bibfield  {author} {\bibinfo {author} {\bibfnamefont {T.}~\bibnamefont
  {Rakovszky}}, \bibinfo {author} {\bibfnamefont {F.}~\bibnamefont
  {Pollmann}},\ and\ \bibinfo {author} {\bibfnamefont {C.~W.}\ \bibnamefont
  {von Keyserlingk}},\ }\bibfield  {title} {\bibinfo {title} {Diffusive
  hydrodynamics of out-of-time-ordered correlators with charge conservation},\
  }\href {https://doi.org/10.1103/PhysRevX.8.031058} {\bibfield  {journal}
  {\bibinfo  {journal} {Phys. Rev. X}\ }\textbf {\bibinfo {volume} {8}},\
  \bibinfo {pages} {031058} (\bibinfo {year} {2018})}\BibitemShut {NoStop}%
\bibitem [{\citenamefont {Nahum}\ \emph {et~al.}(2017)\citenamefont {Nahum},
  \citenamefont {Ruhman}, \citenamefont {Vijay},\ and\ \citenamefont
  {Haah}}]{NahumHaah2017}%
  \BibitemOpen
  \bibfield  {author} {\bibinfo {author} {\bibfnamefont {A.}~\bibnamefont
  {Nahum}}, \bibinfo {author} {\bibfnamefont {J.}~\bibnamefont {Ruhman}},
  \bibinfo {author} {\bibfnamefont {S.}~\bibnamefont {Vijay}},\ and\ \bibinfo
  {author} {\bibfnamefont {J.}~\bibnamefont {Haah}},\ }\bibfield  {title}
  {\bibinfo {title} {Quantum entanglement growth under random unitary
  dynamics},\ }\href {https://doi.org/10.1103/PhysRevX.7.031016} {\bibfield
  {journal} {\bibinfo  {journal} {Phys. Rev. X}\ }\textbf {\bibinfo {volume}
  {7}},\ \bibinfo {pages} {031016} (\bibinfo {year} {2017})}\BibitemShut
  {NoStop}%
\bibitem [{\citenamefont {Nahum}\ \emph {et~al.}(2018)\citenamefont {Nahum},
  \citenamefont {Vijay},\ and\ \citenamefont {Haah}}]{NahumHaah2018}%
  \BibitemOpen
  \bibfield  {author} {\bibinfo {author} {\bibfnamefont {A.}~\bibnamefont
  {Nahum}}, \bibinfo {author} {\bibfnamefont {S.}~\bibnamefont {Vijay}},\ and\
  \bibinfo {author} {\bibfnamefont {J.}~\bibnamefont {Haah}},\ }\bibfield
  {title} {\bibinfo {title} {Operator spreading in random unitary circuits},\
  }\href {https://doi.org/10.1103/PhysRevX.8.021014} {\bibfield  {journal}
  {\bibinfo  {journal} {Phys. Rev. X}\ }\textbf {\bibinfo {volume} {8}},\
  \bibinfo {pages} {021014} (\bibinfo {year} {2018})}\BibitemShut {NoStop}%
\bibitem [{\citenamefont {von Keyserlingk}\ \emph {et~al.}(2018)\citenamefont
  {von Keyserlingk}, \citenamefont {Rakovszky}, \citenamefont {Pollmann},\ and\
  \citenamefont {Sondhi}}]{KeyserlingkSondhi2018}%
  \BibitemOpen
  \bibfield  {author} {\bibinfo {author} {\bibfnamefont {C.~W.}\ \bibnamefont
  {von Keyserlingk}}, \bibinfo {author} {\bibfnamefont {T.}~\bibnamefont
  {Rakovszky}}, \bibinfo {author} {\bibfnamefont {F.}~\bibnamefont
  {Pollmann}},\ and\ \bibinfo {author} {\bibfnamefont {S.~L.}\ \bibnamefont
  {Sondhi}},\ }\bibfield  {title} {\bibinfo {title} {Operator hydrodynamics,
  otocs, and entanglement growth in systems without conservation laws},\ }\href
  {https://doi.org/10.1103/PhysRevX.8.021013} {\bibfield  {journal} {\bibinfo
  {journal} {Phys. Rev. X}\ }\textbf {\bibinfo {volume} {8}},\ \bibinfo {pages}
  {021013} (\bibinfo {year} {2018})}\BibitemShut {NoStop}%
\bibitem [{\citenamefont {Khemani}\ \emph {et~al.}(2018)\citenamefont
  {Khemani}, \citenamefont {Vishwanath},\ and\ \citenamefont
  {Huse}}]{KhemaniHuse2018}%
  \BibitemOpen
  \bibfield  {author} {\bibinfo {author} {\bibfnamefont {V.}~\bibnamefont
  {Khemani}}, \bibinfo {author} {\bibfnamefont {A.}~\bibnamefont
  {Vishwanath}},\ and\ \bibinfo {author} {\bibfnamefont {D.~A.}\ \bibnamefont
  {Huse}},\ }\bibfield  {title} {\bibinfo {title} {Operator spreading and the
  emergence of dissipative hydrodynamics under unitary evolution with
  conservation laws},\ }\href {https://doi.org/10.1103/PhysRevX.8.031057}
  {\bibfield  {journal} {\bibinfo  {journal} {Phys. Rev. X}\ }\textbf {\bibinfo
  {volume} {8}},\ \bibinfo {pages} {031057} (\bibinfo {year}
  {2018})}\BibitemShut {NoStop}%
\bibitem [{\citenamefont {Balachandran}\ \emph {et~al.}(2021)\citenamefont
  {Balachandran}, \citenamefont {Benenti}, \citenamefont {Casati},\ and\
  \citenamefont {Poletti}}]{Balachandran2021}%
  \BibitemOpen
  \bibfield  {author} {\bibinfo {author} {\bibfnamefont {V.}~\bibnamefont
  {Balachandran}}, \bibinfo {author} {\bibfnamefont {G.}~\bibnamefont
  {Benenti}}, \bibinfo {author} {\bibfnamefont {G.}~\bibnamefont {Casati}},\
  and\ \bibinfo {author} {\bibfnamefont {D.}~\bibnamefont {Poletti}},\
  }\bibfield  {title} {\bibinfo {title} {From the eigenstate thermalization
  hypothesis to algebraic relaxation of {OTOCs} in systems with conserved
  quantities},\ }\bibfield  {journal} {\bibinfo  {journal} {Physical Review B}\
  }\textbf {\bibinfo {volume} {104}},\ \href
  {https://doi.org/10.1103/physrevb.104.104306} {10.1103/physrevb.104.104306}
  (\bibinfo {year} {2021})\BibitemShut {NoStop}%
\bibitem [{\citenamefont {Balachandran}\ \emph {et~al.}(2022)\citenamefont
  {Balachandran}, \citenamefont {Santos}, \citenamefont {Rigol},\ and\
  \citenamefont {Poletti}}]{Balachandran2022}%
  \BibitemOpen
  \bibfield  {author} {\bibinfo {author} {\bibfnamefont {V.}~\bibnamefont
  {Balachandran}}, \bibinfo {author} {\bibfnamefont {L.~F.}\ \bibnamefont
  {Santos}}, \bibinfo {author} {\bibfnamefont {M.}~\bibnamefont {Rigol}},\ and\
  \bibinfo {author} {\bibfnamefont {D.}~\bibnamefont {Poletti}},\ }\href@noop
  {} {\bibinfo {title} {Effect of symmetries in out-of-time ordered correlators
  in interacting integrable and nonintegrable many-body quantum systems}}
  (\bibinfo {year} {2022}),\ \Eprint {https://arxiv.org/abs/2211.07073}
  {arXiv:2211.07073 [cond-mat.stat-mech]} \BibitemShut {NoStop}%
\bibitem [{\citenamefont {{Lieb}}\ and\ \citenamefont
  {{Robinson}}(1972)}]{LiebRobinson1972}%
  \BibitemOpen
  \bibfield  {author} {\bibinfo {author} {\bibfnamefont {E.~H.}\ \bibnamefont
  {{Lieb}}}\ and\ \bibinfo {author} {\bibfnamefont {D.~W.}\ \bibnamefont
  {{Robinson}}},\ }\bibfield  {title} {\bibinfo {title} {{The finite group
  velocity of quantum spin systems}},\ }\href
  {https://doi.org/10.1007/BF01645779} {\bibfield  {journal} {\bibinfo
  {journal} {Communications in Mathematical Physics}\ }\textbf {\bibinfo
  {volume} {28}},\ \bibinfo {pages} {251} (\bibinfo {year} {1972})}\BibitemShut
  {NoStop}%
\bibitem [{\citenamefont {Srednicki}(1996)}]{Srednicki_1996}%
  \BibitemOpen
  \bibfield  {author} {\bibinfo {author} {\bibfnamefont {M.}~\bibnamefont
  {Srednicki}},\ }\bibfield  {title} {\bibinfo {title} {Thermal fluctuations in
  quantized chaotic systems},\ }\href
  {https://doi.org/10.1088/0305-4470/29/4/003} {\bibfield  {journal} {\bibinfo
  {journal} {Journal of Physics A: Mathematical and General}\ }\textbf
  {\bibinfo {volume} {29}},\ \bibinfo {pages} {L75–L79} (\bibinfo {year}
  {1996})}\BibitemShut {NoStop}%
\bibitem [{\citenamefont {Deutsch}(1991)}]{Deutsch}%
  \BibitemOpen
  \bibfield  {author} {\bibinfo {author} {\bibfnamefont {J.~M.}\ \bibnamefont
  {Deutsch}},\ }\bibfield  {title} {\bibinfo {title} {Quantum statistical
  mechanics in a closed system},\ }\href
  {https://doi.org/10.1103/PhysRevA.43.2046} {\bibfield  {journal} {\bibinfo
  {journal} {Phys. Rev. A}\ }\textbf {\bibinfo {volume} {43}},\ \bibinfo
  {pages} {2046} (\bibinfo {year} {1991})}\BibitemShut {NoStop}%
\bibitem [{\citenamefont {Srednicki}(1999)}]{Srednicki_1998}%
  \BibitemOpen
  \bibfield  {author} {\bibinfo {author} {\bibfnamefont {M.}~\bibnamefont
  {Srednicki}},\ }\bibfield  {title} {\bibinfo {title} {The approach to thermal
  equilibrium in quantized chaotic systems},\ }\href
  {https://doi.org/10.1088/0305-4470/32/7/007} {\bibfield  {journal} {\bibinfo
  {journal} {Journal of Physics A: Mathematical and General}\ }\textbf
  {\bibinfo {volume} {32}},\ \bibinfo {pages} {1163} (\bibinfo {year}
  {1999})}\BibitemShut {NoStop}%
\bibitem [{\citenamefont {Huang}\ \emph {et~al.}(2019)\citenamefont {Huang},
  \citenamefont {Brand\~ao},\ and\ \citenamefont {Zhang}}]{HuangZhang2019}%
  \BibitemOpen
  \bibfield  {author} {\bibinfo {author} {\bibfnamefont {Y.}~\bibnamefont
  {Huang}}, \bibinfo {author} {\bibfnamefont {F.~G. S.~L.}\ \bibnamefont
  {Brand\~ao}},\ and\ \bibinfo {author} {\bibfnamefont {Y.-L.}\ \bibnamefont
  {Zhang}},\ }\bibfield  {title} {\bibinfo {title} {Finite-size scaling of
  out-of-time-ordered correlators at late times},\ }\href
  {https://doi.org/10.1103/PhysRevLett.123.010601} {\bibfield  {journal}
  {\bibinfo  {journal} {Phys. Rev. Lett.}\ }\textbf {\bibinfo {volume} {123}},\
  \bibinfo {pages} {010601} (\bibinfo {year} {2019})}\BibitemShut {NoStop}%
\bibitem [{\citenamefont {Cheneau}\ \emph {et~al.}(2012)\citenamefont
  {Cheneau}, \citenamefont {Barmettler}, \citenamefont {Poletti}, \citenamefont
  {Endres}, \citenamefont {Schau{\ss}}, \citenamefont {Fukuhara}, \citenamefont
  {Gross}, \citenamefont {Bloch}, \citenamefont {Kollath},\ and\ \citenamefont
  {Kuhr}}]{CheneauKuhr2012}%
  \BibitemOpen
  \bibfield  {author} {\bibinfo {author} {\bibfnamefont {M.}~\bibnamefont
  {Cheneau}}, \bibinfo {author} {\bibfnamefont {P.}~\bibnamefont {Barmettler}},
  \bibinfo {author} {\bibfnamefont {D.}~\bibnamefont {Poletti}}, \bibinfo
  {author} {\bibfnamefont {M.}~\bibnamefont {Endres}}, \bibinfo {author}
  {\bibfnamefont {P.}~\bibnamefont {Schau{\ss}}}, \bibinfo {author}
  {\bibfnamefont {T.}~\bibnamefont {Fukuhara}}, \bibinfo {author}
  {\bibfnamefont {C.}~\bibnamefont {Gross}}, \bibinfo {author} {\bibfnamefont
  {I.}~\bibnamefont {Bloch}}, \bibinfo {author} {\bibfnamefont
  {C.}~\bibnamefont {Kollath}},\ and\ \bibinfo {author} {\bibfnamefont
  {S.}~\bibnamefont {Kuhr}},\ }\bibfield  {title} {\bibinfo {title}
  {{Light-cone-like spreading of correlations in a quantum many-body system}},\
  }\href {https://doi.org/10.1038/nature10748} {\bibfield  {journal} {\bibinfo
  {journal} {{Nature}}\ }\textbf {\bibinfo {volume} {481}},\ \bibinfo {pages}
  {484 } (\bibinfo {year} {2012})}\BibitemShut {NoStop}%
\bibitem [{\citenamefont {Luitz}\ \emph {et~al.}(2020)\citenamefont {Luitz},
  \citenamefont {Moessner}, \citenamefont {Sondhi},\ and\ \citenamefont
  {Khemani}}]{LuitzKhemani}%
  \BibitemOpen
  \bibfield  {author} {\bibinfo {author} {\bibfnamefont {D.~J.}\ \bibnamefont
  {Luitz}}, \bibinfo {author} {\bibfnamefont {R.}~\bibnamefont {Moessner}},
  \bibinfo {author} {\bibfnamefont {S.~L.}\ \bibnamefont {Sondhi}},\ and\
  \bibinfo {author} {\bibfnamefont {V.}~\bibnamefont {Khemani}},\ }\bibfield
  {title} {\bibinfo {title} {Prethermalization without temperature},\ }\href
  {https://doi.org/10.1103/PhysRevX.10.021046} {\bibfield  {journal} {\bibinfo
  {journal} {Phys. Rev. X}\ }\textbf {\bibinfo {volume} {10}},\ \bibinfo
  {pages} {021046} (\bibinfo {year} {2020})}\BibitemShut {NoStop}%
\bibitem [{\citenamefont {Lee}\ \emph {et~al.}(2019)\citenamefont {Lee},
  \citenamefont {Kim},\ and\ \citenamefont {Kim}}]{Lee2019}%
  \BibitemOpen
  \bibfield  {author} {\bibinfo {author} {\bibfnamefont {J.}~\bibnamefont
  {Lee}}, \bibinfo {author} {\bibfnamefont {D.}~\bibnamefont {Kim}},\ and\
  \bibinfo {author} {\bibfnamefont {D.-H.}\ \bibnamefont {Kim}},\ }\bibfield
  {title} {\bibinfo {title} {Typical growth behavior of the out-of-time-ordered
  commutator in many-body localized systems},\ }\href
  {https://doi.org/10.1103/PhysRevB.99.184202} {\bibfield  {journal} {\bibinfo
  {journal} {Phys. Rev. B}\ }\textbf {\bibinfo {volume} {99}},\ \bibinfo
  {pages} {184202} (\bibinfo {year} {2019})}\BibitemShut {NoStop}%
\bibitem [{\citenamefont {{Bohigas}}\ \emph {et~al.}(1984)\citenamefont
  {{Bohigas}}, \citenamefont {{Giannoni}},\ and\ \citenamefont
  {{Schmit}}}]{BGS1984}%
  \BibitemOpen
  \bibfield  {author} {\bibinfo {author} {\bibfnamefont {O.}~\bibnamefont
  {{Bohigas}}}, \bibinfo {author} {\bibfnamefont {M.~J.}\ \bibnamefont
  {{Giannoni}}},\ and\ \bibinfo {author} {\bibfnamefont {C.}~\bibnamefont
  {{Schmit}}},\ }\bibfield  {title} {\bibinfo {title} {{Characterization of
  Chaotic Quantum Spectra and Universality of Level Fluctuation Laws}},\ }\href
  {https://doi.org/10.1103/PhysRevLett.52.1} {\bibfield  {journal} {\bibinfo
  {journal} {\prl}\ }\textbf {\bibinfo {volume} {52}},\ \bibinfo {pages} {1}
  (\bibinfo {year} {1984})}\BibitemShut {NoStop}%
\bibitem [{\citenamefont {Casati}\ \emph {et~al.}(1980)\citenamefont {Casati},
  \citenamefont {Valz-Gris},\ and\ \citenamefont {Guarnieri}}]{Casati1980}%
  \BibitemOpen
  \bibfield  {author} {\bibinfo {author} {\bibfnamefont {G.}~\bibnamefont
  {Casati}}, \bibinfo {author} {\bibfnamefont {F.}~\bibnamefont {Valz-Gris}},\
  and\ \bibinfo {author} {\bibfnamefont {I.}~\bibnamefont {Guarnieri}},\
  }\bibfield  {title} {\bibinfo {title} {Connection between quantization of
  nonintegrable systems and statistical theory of spectra},\ }\href@noop {}
  {\bibfield  {journal} {\bibinfo  {journal} {Lett. Nuovo Cimento}\ }\textbf
  {\bibinfo {volume} {28}},\ \bibinfo {pages} {279} (\bibinfo {year}
  {1980})}\BibitemShut {NoStop}%
\bibitem [{\citenamefont {Oganesyan}\ and\ \citenamefont
  {Huse}(2007)}]{OganesyanHuse2007}%
  \BibitemOpen
  \bibfield  {author} {\bibinfo {author} {\bibfnamefont {V.}~\bibnamefont
  {Oganesyan}}\ and\ \bibinfo {author} {\bibfnamefont {D.~A.}\ \bibnamefont
  {Huse}},\ }\bibfield  {title} {\bibinfo {title} {Localization of interacting
  fermions at high temperature},\ }\href
  {https://doi.org/10.1103/PhysRevB.75.155111} {\bibfield  {journal} {\bibinfo
  {journal} {Phys. Rev. B}\ }\textbf {\bibinfo {volume} {75}},\ \bibinfo
  {pages} {155111} (\bibinfo {year} {2007})}\BibitemShut {NoStop}%
\bibitem [{NSC()}]{NSCC}%
  \BibitemOpen
  \href@noop {} {}\bibinfo {howpublished} {https://www.nscc.sg/, access from
  January 1st to September 30th 2022.}\BibitemShut {Stop}%
\end{thebibliography}%

\end{document}